# Chapter 1

# Edge Probes of Topological Order

Moty Heiblum[1] and D. E. Feldman[2]

[1]Department of Condensed Matter Physics, Weizmann Institute of Science, Rehovot, Israel. [2]Brown Theoretical Physics Center and Department of Physics, Brown University, Providence, RI 02912-1843, USA

**Abstract**

According to the *bulk-edge* correspondence principle, the physics of the gapless edge in the quantum Hall effect determines topological order in the gapped bulk. As the bulk is less accessible, the last two decades saw the emergence of several experimental techniques that invoke the study of the compressible edge. We review the properties of the edge, and describe several experimental techniques that include shot noise and thermal noise measurements, interferometry, and energy (thermal) transport at the edge. We pay special attention to the filling factor 5/2 in the first excited Landau level (in two-dimensional electron gas in GaAs), where experimental evidence of a non-abelian topological order was found. A brief discussion is devoted to recent interferometry experiments that uncovered unexpected physics in the integer quantum Hall effect. The chapter also addresses the theory of edge states, for systems with abelian and non-abelian topological orders.

## 1. Introduction

The fractional quantum Hall effect (FQHE) famously gives rise to quasiparticles whose charges are lower than an electron charge. Their exchange statistics is neither Fermi nor Bose. Roughly speaking, the knowledge of the topological order in FQHE systems consists in the understanding of the charges and the statistics of those fractionally charged *anyons*. Since the anyons exist in the gapped bulk of 2D electron liquid, experiments probing the bulk may a priori seem the best way to test the order of the state. Yet, such probes, being difficult to realize, are often not useful in respect to the latter. Hence, the focus of this chapter is on edge physics. The electron liquid is gapless at the sample's edges, where the filling factor changes from the bulk value to zero, and the bulk topological order, protected by the bulk gap, is absent. Naively, this makes the edge a wrong place to look for signatures of the topological order. Nevertheless, most of our knowledge of the topological order in the FQHE regime comes from probing the edge.

The most basic question involves the charges of the *anyons*. Edge physics can shed light on this question, e.g., through measuring *shot noise*. Weak tunneling between opposite edges of the sample, brought to nearby proximity by means of a narrow constriction, leads to a universal noise that is proportional to the carriers' charge. Additional information comes from strong tunneling experiments. This physics is reviewed in Section 3 below.

Since the exchange statistics of *anyons* is defined in terms of particles moving around each other, *anyonic* interferometry is a natural tool to understand the states' topological orders. In turn, interferometry is studied by splitting the incident chiral current emanating from the source into two (or more) different paths, and merging them again before the drain. We address interferometry in Section 5, focusing on the integer QHE (IQHE). Understanding interferometry in the IQHE regime is mandatory before one can rely on its interpretation in the FQHE. Recent developments, such as the discovery of the *pairing effect* and the *lobe structure*, underscore how challenging interferometry is even in the IQHE.

Another approach has recently emerged as a powerful probe of topological orders: experiments involving energy transport. We first address the physics of *neutral modes* in Section 4. The simplest quantum Hall edges carry only *downstream* charged channels (the chirality is dictated by the magnetic field). However, more typically, charged channels may coexists with one or more (often *upstream*, namely, with opposite chirality) *neutral* modes. Since the neutral modes (not carrying net charge) weakly interact with external probes, accessing them is a difficult task. A breakthrough in that direction came in 2010, when the existence of *upstream neutral* modes was proven experimentally via shot noise measurements. An even more important development consisted in the measurements of the thermal conductance of quantum Hall edge modes (Section 7). In particular, a thermal conductance experiment gave strong support to *non-abelian statistics* at filling factor $\nu = 5/2$.

Section 6 focuses on that enigmatic $\nu = 5/2$ filling factor. Its very existence was a surprise. The most intuitive way to think of FQHE states is based on *composite fermions* (CFs). Odd-denominator fractions can be interpreted as IQHE gapped states of non-interacting CFs. On the other hand, a picture of weakly interacting CFs suggests *gapless* states at even-denominator filling factors. This agrees with the experiments in the first two spin-split Landau levels (LLs), but not in higher LLs. The apparent solution of this paradox lies in a picture of Cooper-pairing of CFs. Yet, the solution gives rise to another conundrum: numerous ways exist to build Cooper pairs and instead of 'no obvious candidates for the $\nu = 5/2$ topological liquid' we now have way too many. Until new experimental probes of neutral modes were developed, the scant amount of available experimental data shifted the focus to numerical calculations. Different orders were leading at different times, and most recently numerics gave support to the *Pfaffian* and *anti-Pfaffian* topological orders, exhibiting the beautiful property of non-abelian statistics. This feature is of current great interest for quantum computing.

A very recent thermal conductance experiment does support a non-abelian liquid at $\nu = 5/2$; however the topological order appears to be neither of the two leading numerical candidates. The best fit to the existing data comes from the *PH-Pfaffian* topological order. Presumably, disorder effects, neglected in numerics, may be responsible for this order. More work is needed until the order of the $\nu = 5/2$ liquid and its universality are settled.

In addition to key Sections 3-7 and a summary, this chapter reviews some background information: edge models in Section 2 and in Appendix A, *bulk-edge* correspondence at $\nu = 5/2$ in Appendix B, and Coulomb effects in interferometry in Appendix C.

## 2. Edge channels

Halperin was the first to suggest that edge-channels transport is responsible for the conduction mechanism in the QHE [1]. According to this successful model, current flows along the edges of the sample, with electrons performing classical chiral 'skipping orbits'. In the quantum mechanical language, current flows at the crossing of the Landau levels and the Fermi energy near the physical edges of the sample. In the IQHE regime, electron correlations are weak, and currents flow in non-interacting 1D-like channels. Alternatively, in the FQHE regime, electron correlations are strong, leading the edge channels to possess chiral Luttinger liquid-like properties [2]. Due to the chirality, backscattering in wide samples is minimized and the edge channels tend to propagate 'ballistically' for a long distance [3]. Since the interpretation of the experiments discussed here depends on the inner structure of the edge channels, we start with a brief introduction to edge channel models (excellent discussions can be found in the reviews by Kane and Fisher [4] and by Wen [3]).

We begin with the simplest spin-polarized non-interacting state with filling factor $\nu = 1$ (the lowest spin-split Landau level). Our starting point is a two-dimensional electron gas (2DEG) in an external potential $V(y)$, with an edge at $y = 0$ and a perpendicular magnetic field in the $z$-direction (Fig. 1).

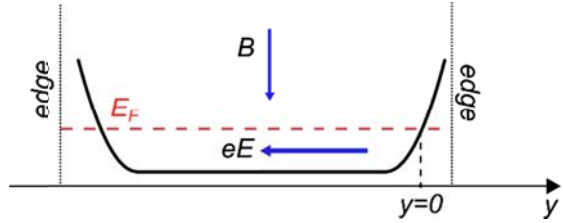

Fig. 1. Electrons near the edge at $y = 0$ experience a perpendicular magnetic field $B$ in the $z$ direction, and an electrostatic force $eE$ due to the confining potential in the $y$ direction. The thick line shows a simplified confining potential.

The system is infinite in the $x$-direction. In the Landau gauge, $A_x = By$, $A_y = A_z = 0$, the infinitely degenerate wave-functions of the lowest Landau level are

$$\psi_k(x, y) = exp\left(ikx - \frac{[y-y_0]^2}{2l_B^2}\right) \tag{1}$$

where $y_0 = c\hbar k/|e|B$, and the magnetic length $l_B = \sqrt{\hbar c/|e|B}$ ($\sqrt{\hbar/|e|B}$ in the MKS units). These wave-functions describe charge density in narrow strips running along the $x$-direction and with width $\sim l_B$ along the $y$ direction. When the potential $V(y)$ changes on a much larger scale than $l_B$, it does not affect the wave-functions but lifts their degeneracy. Only the states below the Fermi energy $E_F = \frac{\hbar|e|B}{mc} + V(y = 0)$ are filled, that is, the states with $y_0 < 0$ (corresponding to a positive edge velocity). Hence, our model describes an edge channel at $y = 0$ with a small width, where the charge density changes from zero at the edge to the bulk value inside the liquid.

We address that limit by linearizing the potential near $y = 0$ and expressing the Hamiltonian as,

$$H = \sum_k vk\psi_k^+\psi_k \tag{2}$$

where the fermion operators create and annihilate electrons in the states $\psi_k(x, y)$. The states are parametrized by the momentum $k$ in the $x$-direction, and the problem reduces effectively to 1D with all the excitations propagating *downstream* with the same drift velocity $v$. The physical origin of the chirality is apparent in Fig. 1, with charges drifting at the edge in perpendicular electric and magnetic fields. This is also evident from the semi-classical picture of skipping orbits [5]. The Hamiltonian $H$ can be seen as a chiral part of a Hamiltonian of a non-chiral quantum wire. A similar Hamiltonian describes the opposite edge of the Hall bar, where excitations move in the opposite direction (at the same chirality). The Landauer conductance $e^2/h$ of a quantum wire agrees with what one expects in the IQHE.

Even though electrons are fermions, it is often convenient to rewrite the edge model in terms of Bose operators. This approach does not make much difference in a non-interacting problem, but has advantages if Coulomb interactions at the edge are taken into account. While in terms of electron edge transport, the FQHE regime is a hopelessly complicated problem of strongly interacting fermions, many of the difficulties magically disappear in the Bose language, providing an easy way to describe the low-energy excitations, which are bosonic plasmons [3].

We focus on the 1D charge density $\rho(x)$ along the edge, obtained by integrating the charge distribution along the $y$ direction. Defining the Bose field $\varphi(x)$ according to $\rho(x) = e\partial_x\varphi(x)/2\pi$, we get an *edge-action*:

$$\frac{S}{\hbar} = -\frac{1}{4\pi}\int dx\, dt[\partial_t\varphi\partial_x\varphi + v(\partial_x\varphi)^2] \qquad (3)$$

which we further briefly discuss in Appendix A. This action is a minimal model of the edge in the IQHE regime, as additional terms are unimportant at low energies. When long-range Coulomb interactions are important, another term, $\int dx_1 dx_2 \rho(x_1)\rho(x_2)/\epsilon|x_1 - x_2|$, must be included. The interaction is responsible particularly for a high edge velocity on an ungated edge and for a 'striped' structure of an edge, which may reconstruct in realistic samples due to Coulomb interaction. As was predicted in[6, 7], and observed in [8], the edge may consist of compressible and incompressible stripes that form multiple parallel conducting channels, resulting from edge-reconstruction [9], thus necessitating the addition to the action of the contributions from pairs of counter-propagating identical channels. Such pairs of channels are subject to Anderson localization via impurities on relatively short distances [10]. An extension of the action in Eq. (3) to integer states with filling $\nu > 1$ is easy: each filled Landau level leads to one edge channel.

## 2.1. *Particle-states in the FQHE regime*

We now turn to the FQHE and discuss the basics of the chiral Luttinger liquid theory, introduced by Wen [3]. For Laughlin states at $\nu = 1/(2p + 1)$, there are only minute changes in the expressions from those in the IQHE. This can be understood from Jain's Composite Fermion (CF) picture [11]. With a CF formed after inserting an even number $2p$ of flux quanta for each electron ($2p\Phi_0 = 2ph/e$), each CF 'feels' a reduced effective magnetic field (in the mean-field approximation). Alternatively, it is legitimate to think of each electron as a combination of $2p$ flux quanta with a single CF [11, 12]. Consequently, at $\nu = 1/2$, the CF is immersed in a zero effective magnetic field. At $\nu = 1/3$, its effective filling factor is 1, and a single edge channel is expected in the above action. This maintains the general structure of Eq. (3), with the only difference being the prefactor due to a different electrical conductance, $\nu e^2/h$ (see Appendix A):

$$\frac{S}{\hbar} = -\frac{1}{4\pi\nu}\int dx\, dt[\partial_t\varphi\partial_x\varphi + v(\partial_x\varphi)^2] \qquad (4)$$

The physics is richer at other fractional filling factors. Consider $\nu = 2/5$ as an example. The CF description with $p=1$ leads to two filled Landau levels of CFs, and thus two *downstream* edge channels. The simplest model is thus a sum of the two actions of the type described in Eqs. (3) and (4), but with added terms representing the Coulomb interaction between the two channels and inter-channel charge tunneling. We postpone a discussion of charge tunneling to Section 4.

### 2.2. *Hole-Conjugate states in the FQHE regime*

The $\nu = 2/3$ state and other states in the range $1/2 < \nu < 1$ (the 'hole-conjugate' states) behave quite differently than the particle-states. The $\nu = 2/3$ liquid can be thought as an FQHE state of $\nu = 1/3$ 'holes' on top of an IQHE electron liquid at $\nu = 1$. Imagine that the charge density looks roughly like depicted in Fig. 2.

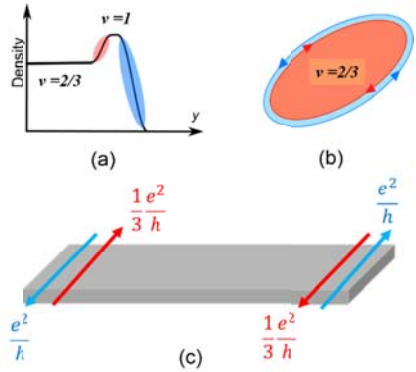

Fig. 2. MacDonald's clean edge configuration, with the edge modes of the ν=2/3 state. (a) Potential (and density) distribution near the edge. (b) Chirality is counter-clockwise, the $\nu = 1$ channel (blue) moves *downstream*, while the $\nu = 1/3$ channel (red) moves *upstream*. (c) Two-terminal conductance $G$=(1+1/3) $e^2$/h. With inter-mode scattering, the *downstream* charge channel has conductance $G$=2$e^2$/3h being accompanied by an *upstream* neutral mode (not shown).

As the picture suggests, there are two counter-propagating edge channels. Since the electrostatic potential follows the electron density, the direction of the confining electric field is opposite at the two interfaces: one separating $\nu = 1$ from $\nu = 0$ (at the sample's edge), and the other separating $\nu = 1$ from $\nu = 2/3$ (equivalently, $\nu = 0$ of holes from $\nu = 1/3$ of holes). Two oppositely propagating channels emerge: a *downstream* channel identical to that of *ν*=1 and an *upstream* channel of $\nu = 1/3$ [13]. On a short scale, inter-channel equilibration does not take place with the theoretically expected (but never observed) two-terminal conductance of $4e^2/3h$ (Fig. 2c) [14]. At longer edges, after charge equilibration, the two-terminal conductance drops to $2e^2/3h$ (as is always experimentally observed; see also later). Edge-reconstruction may lead to additional pairs of counter-propagating modes on the edges of realistic samples. In particular, a pair of modes of conductance $e^2/3h$ was found [15, 16], replacing the modes' picture of MacDonald [13]. A general way to construct edge actions is known as the *K*-matrix formalism [3]. We briefly summarize it in Appendix A.

## 3. Shot noise and charge in the FQHE

### 3.1. *Theoretical considerations*

One of the most striking features of the FQHE is its excitations that carry a fraction of an electron charge. Such excitations are an inevitable consequence of the coexistence of the quantized Hall conductance $\sigma = \nu e^2/h$ with an energy gap in the bulk [17]. Following Laughlin, consider an FQHE system pierced by a solenoid in the middle of a plaquette in a lattice on which electrons move (electrons cannot enter the solenoid and are not affected by it as long as the magnetic field through the solenoid is zero). Next, the magnetic flux slowly increases within the solenoid by one flux quantum $\Phi_0$. The time-dependent flux gives rise to a circular electric field $E$ around the solenoid and its attendant electric current in the bulk away from the solenoid. Charge moves away from the pierced plaquette to the edge of the sample. The total transferred charge is $\Delta q = \sigma \int dt dl E = \nu e$, where the integration extends over a circle inside the 2D electron gas, with the center at the solenoid. The depleted charge is $\nu e$ - being the charge of an excitation. Indeed, at the end of the process, we are left with one flux quantum through the piercing solenoid. Such flux is invisible to electrons in the 2D gas and can be eliminated by a large gauge transformation. Hence, the final charge distribution corresponds to an eigenstate of the initial Hamiltonian without the added flux, and the depleted charge is indeed an excitation charge. Thus, the observed fractional quantized conductance leads, though not in a direct way, to a fractional excitation charge.

Note that it is not necessarily the minimal possible charge. The reason is that the system is made of electrons, and hence, there are always excitations of the charge $e$. For $\nu = m/n$, with co-primes $m$ and $n$, one can always build an excitation of charge $e/n$ from several charges $e$ and $\nu e$. For an even $n$, the minimal charge was argued to be even lower, $e/2n$ [18].

Edge physics can be used for direct evidence of charge fractionalization by measuring shot noise. The idea is a century old and was employed by Schottky in the early 20th century to measure the charge of an electron in vacuum tubes [19-21].

Imagine charge transport through a barrier by rare tunneling events. The current $I = ep$, where $p$ is the tunneling rate, is transporting the total charge $Q = ept$ during the time $t$. Its mean-square-fluctuation is $\langle \Delta Q^2 \rangle = e^2 p t$ and the spectral density, defined as $S_i = 2\langle \Delta Q^2 \rangle / t$, in units of A$^2$/Hz is equal to:

$$S_i = \int dt \langle I(t)I(0) + I(0)I(t) \rangle = 2eI \qquad (5)$$

where we substituted $Q = \int I(t)dt$. Thus, measurements of the noise and the current yield the carrier's charge. Nothing above implies that the carriers must be electrons. If the current comes from rare tunneling events of charge $e^*$ quasiparticles, then $S_i = 2e^*I$ (Fig. 3).

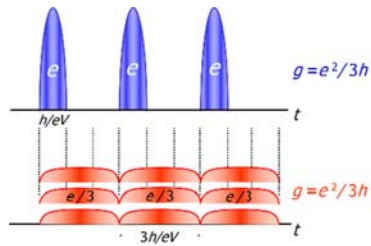

Fig. 3. Trains of electrons (top) and fractional charges (bottom), with both leading to the same conductance. After partitioning of charges; e.g., with a partly pinched QPC, the shot noise is proportional to the particle charge.

Experimentally, a quantum point contact (QPC) constriction brings two opposite edge channels to nearby proximity, allowing charge to tunnel between the channels. The minimal tunneling charge is set by the bulk, being the charge of a quasiparticle [22, 23]. We thus turn to shot noise experiments in FQHE.

**3.2.** *Measurements of shot noise*

Shot noise in mesoscopic conductors has proved to be a powerful experimental probe [19, 22-24]. The noise provides information that is not necessarily available from the time-average flux of the particles (see Fig. 3), such as the particles' charge, their correlations, and even their temperature.

At zero temperature, an unpartitioned edge channel is noiseless - a direct consequence of Fermi statistics [24-26]. At a finite temperature, the channel carries also thermal noise, which is a property of any conductor, independent of its microscopic details and the carriers' charge. The spectral density of the added noise takes the form $S_T=4k_B Tg$, with $k_B$ the Boltzmann's constant, $T$ the temperature, and $g$ the conductance of the edge channel. The so-called 'quantum shot noise' (the *excess noise* above the thermal noise) differs from the classical shot noise by reflecting the noise-free property of the emitting reservoir and the finite tunneling strength [24-26]. This was demonstrated first in the simplest mesoscopic system - a QPC [27, 28]. At zero temperature, the contribution of the partitioned $p$'th channel (on a multi-channel edge), and a weakly back-scattering QPC ($t_p\sim 1$) is

$$S_i(0) = 2e^* V g_p t_p (1-t_p) , \qquad (6)$$

where $S_i(0)$ is the 'zero frequency' spectral density ($f<<e^*V/h$), $V$ the applied source voltage, $g_p$ the conductance of the fully transmitted $p$'th channel, $e^*$ the quasiparticle charge, and $t_p$ the effective transmission coefficient of the $p$'th channel (with $I_p=Vg_p t_p$) in the QPC [24, 29]. Note, that Eq. (6) can also be used when back-scattering is strong ($t_p<<1$), but with $e^*$ replaced by $e$ [22, 23]. The first limit in Eq. (6) corresponds to a constriction between two channels with the FQHE bulk filling, while the second limit emerges when the two channels are separated by a $\nu\sim 0$ insulating region (a nearly depleted constriction). Clearly, only electrons are allowed to tunnel in the second case.

To make the process of charge determination clearer, the fully reflected inner channels and fully transmitted outer ones do not contribute to the shot noise; only the partitioned channel does (Fig. 4). For example, at $\nu = 2/5$ with $p$=2, consider the excess noise generated by partitioning the inner CF channel, which separates $\nu = 2/5$ from $\nu = 1/3$. The effective transmission of the inner channel is $g = (g - g_{1/3})/(g_{2/5} - g_{1/3})$, with the transmitted current it carries being $I_t = V(g_{2/5} - g_{1/3})t_{\text{inner}}$, where $g$ is the actual conductance of the partly pinched QPC, while $g_{2/5}$ and $g_{1/3}$ are the corresponding quantum conductances of the fractional states.

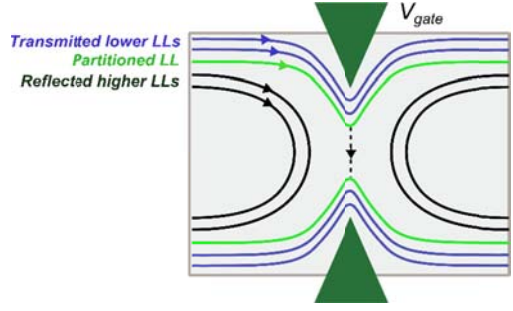

Fig. 4. Partitioning of edge modes by a partly pinched QPC.

At a finite temperature, the expression for the spectral density of non-interacting fermions of charge $e^*$ takes the form [24],

$$S = 2e^* V g_p t_p (1-t_p) \left[ \coth\left(\frac{e^* V}{2k_B T}\right) - \frac{2k_B T}{e^* V} \right] + 4k_B T g \qquad (7)$$

Since the tunneling particles are interacting fermions, the choice of Eq. (7) to fit the FQHE data may seem unjustified. We justify it at the end of this Section. At any rate, when V>>VT ~2kBT/e*, the noise approaches a linear dependence on V (and I), as predicted by the zero temperature expression (Eq. (6)). It should be noted that different approaches to determine the quasiparticles charge were employed. They include, for example, tunneling with a single electron transistor to a localized puddle of quasiparticles [30], or resonant tunneling of quasiparticles into an isolated island (or an impurity) [31-33].

### 3.3. *Experimental techniques*

Heterostructures, hosting 2DEG with typical carrier density $n_s$=(1-2)·$10^{11}$cm$^{-2}$ and mobility $\mu$=(2-30)·$10^6$cm$^2$/V-s at 0.3K, were employed for a variety of measurements conducted in a dilution refrigerator with electrons' temperature of 10-50mK. Achieving the electron temperature close to the lattice temperature was aided by 'cold grounding' of most of the ohmic contacts (by connecting the contacts with short wires to the 'cold finger'). The noise was filtered by an *LC* circuit with the center frequency ~1MHz and bandwidth 30-100kHz (depending on the sample's resistance). Reflected and transmitted currents from the QPC were collected by different terminals. Voltage fluctuations were amplified by a low-noise homemade preamplifier, cooled to 4.2K. Its typical voltage noise is less than $2.5 \times 10^{-19}$V$^2$/Hz and the current noise ~$10^{-28}$A$^2$/Hz. The signal was amplified by a room temperature amplifier and measured by a spectrum analyzer. Since the accurate magnitude of the noise signal is important, careful calibration of the total gain is routinely performed by measuring a known charge or temperature [34, 35].

### 3.4. *Weak versus strong backscattering*

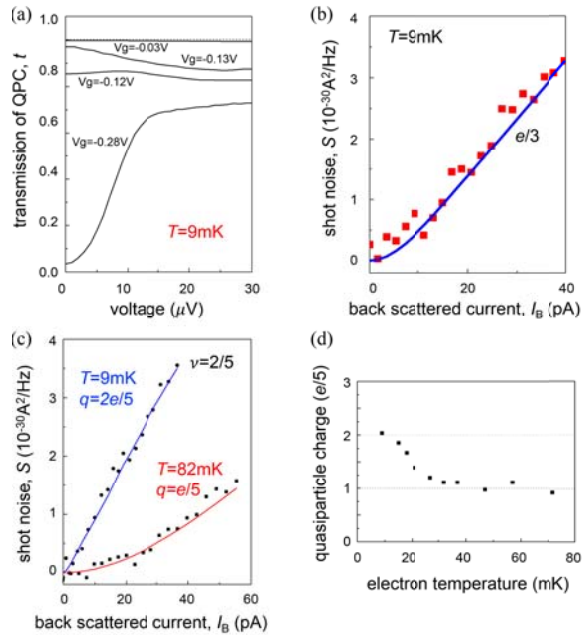

Fig. 5. The non-linear transmission and the excess shot noise in a QPC for fractional fillings $\nu = 1/3$ and $\nu = 2/5$. (a) $\nu = 1/3$. Dependence of the transmission on the applied voltage. Drastic disagreement weak back-scattering regime with the chiral Luttinger liquid prediction. A better agreement with CLL at strong backscattering. (b) $\nu = 1/3$. Excess shot noise measured at extremely weak backscattering, ~0.01 (see in (a) a constant transmission. Note the extremely small excess noise). (c, d) $\nu = 2/5$. Noise measurements at very weak backscattering at different temperatures. The measured charge increases to $e^* = 2e/5$ at the lowest temperature (from Ref. [36]).

We start with noise measurements of a single composite fermion channel at bulk filling $\nu=1/3$ and $g_{1/3}=e^2/3h$. When we use a QPC constriction, even a relatively weak backscattering potential may lead, at the lowest temperature (10-20mK), to a highly non-linear transmission coefficient with a non-universal dependence on the source voltage [36-39] (not agreeing with the chiral Luttinger liquid model [40-43]). Such non-linearity leads to a non-universal excess noise with a larger Fano factor, $F=e^*/e >1/3$. Only at extremely weak backscattering, the transmission coefficient is nearly independent of the source voltage, with the Fano factor $F=e^*/e =1/3$ [36] (Fig. 5).

What about the excess noise at $\nu=2/5$? With nearly unity effective transmission, $t_{inner} = \dfrac{g - g_{1/3}}{g_{2/5} - g_{1/3}}$, the Fano factor increases smoothly with lowering the temperature; being $F=2/5$ at $T\sim9$mK and $F=1/5$ at $T\sim82$mK (Figs. 5a & 5b). A similar trend of the Fano factor is found also at $\nu=3/7$ [36].

In the strong backscattering limit, the quasiparticle charge is expected to approach the electron charge. An example of the measured shot noise at $\nu=1/3$ as a function of the transmission is summarized in Fig. 6, where measurements taken with a few samples collapse onto one curve [44].

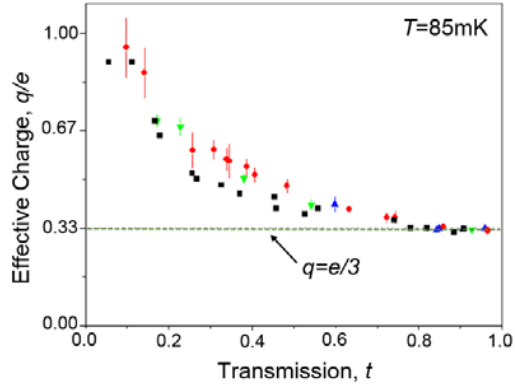

Fig. 6. Evolution of quasiparticle charge as a function of the backscattering strength at $\nu = 1/3$. At weak backscattering strengths the measured charge is $e^*=e/3$, increasing as the QPC pinches, approaching the charge of an electron. Different colors represent measurements in different devices (from Ref. [44]).

### 3.5. *The puzzle of the noise formula*

One piece of the shot noise story has been left out: What is the justification for the fitting formulae in Eqs. (6) and (7)? Empirically, the justification is that they work. A microscopic approach to the noise builds on the chiral Luttinger liquid model with a point impurity [22, 23, 40-43], which allows an exact solution for the I-V characteristic and the shot noise. However, the solution does not agree with the data obtained with a partitioning QPC. While this is not a surprise, as the simplest integrable model misses much of the edge physics (Section 2), the surprise is that simplistic Eqs. (6,7) work.

The key to this success is related to the Johnson-Nyquist (J-N) noise $4k_\text{B}Tg$ in Eq. (7) - a consequence of the general fluctuation-dissipation theorem (FDT). While the J-N formula applies only in thermal equilibrium, its non-equilibrium generalization was discovered from the exact expression for the noise in the integrable model [22, 23, 40-43], as well as in non-integrable models [45-47].

The equilibrium FDT is based on two ingredients: the linear response theory and the Gibbs distribution. Away from equilibrium, only the first ingredient remains; hence, the FDT crumbles. However, in chiral systems the linear response theory is based on the causality principle in the temporal and space domains; namely, any events have consequences only *downstream*. This stronger causality opens a way to derive rigorously numerous model-independent FDT-like theorems for quantum Hall edge channels [45-47]. At weak backscattering, the FDTs agrees quite well with Eq. (7), as long as the transmission is weakly non-linear; yet, Eq. (7) has been also successful for relatively strong backscattering (with $e$ replacing $e^*$ for QPC near pinch-off). This is harder to address since no model-independent equation could be derived in that regime, and only a crude, approximate expression exists [48]. Its analytical structure is rather different from Eq. (7), but amazingly, it is almost numerically indistinguishable from Eq. (7), thus explaining its success. Yet, it also shows that the determined charge $e^*$ can only be interpreted as the charge of the tunneling carriers (electrons or quasiparticles) for transmissions near zero or near one. In other cases, $e^*$ might be seen as some average of multiple charges, involved in transport.

As a simplistic illustration of the latter, imagine, for example, that both quasiparticle and electron tunneling is possible. In a model with rare independent tunneling events of the two carrier types, the current $I = e^*p^* + ep$, where $p$ and $p^*$ are the tunneling rates. The noise $S = 2(e^*)^2 p^* + 2e^2 p = 2qI$, where the effective charge $q = \frac{(e^*)^2 p^* + e^2 p}{e^* p^* + ep}$ is between the charges of an electron and a quasiparticle.

## 4. Neutral Modes

### 4.1. *Theory*

The edge models of the states at fillings $\nu = 1$, 1/3, 2/5, etc. are consistent with the quantization of the bulk conductance (Section 2); however, the simplest model of the $\nu$=2/3 state is not. The most basic edge model includes two non-interacting, counter-propagating channels of the *downstream* conductance $e^2/h$ and *upstream* conductance $e^2/3h$. As discussed below, such a model leads to an incorrect conductance of $4e^2/3h$ (Fig. 2). The correct (measured) two-terminal conductance of $2e^2/3h$ is restored by edge equilibration processes.

The simplest model misses density-density interaction of the charges in the two counter-propagating channels. Yet, such interaction alone cannot change the incorrect value of the conductance. Indeed, in the thermodynamic limit, the total charge of each of the four edge channels on both sides of the Hall bar must be conserved separately (Fig. 2c). Thus, one can assign separate electrochemical potentials to each of the channels. For the two channels, emanating from the grounded terminal, the electrochemical potential is zero, and for the other two channels, emanating from the source, it is $eV$. Charge conservation dictates that the current must be conserved in all the points along the edge, independently of the interaction strength along the edge. This yields incorrect conductance, and hence, something beyond Coulomb effects is needed to solve our problem.

The solution lies in inter-channel tunneling [49, 50], which equilibrates the electrochemical potentials of the counter-propagating channels on each edge. We briefly review the edge theory in the presence of the inter-channel tunneling in Appendix A. The discussion reveals a subtlety: momentum mismatch between the two edge channels makes tunneling ineffective in equilibrating the channels in a clean sample. Thus, the quantized electrical conductance crucially depends on the presence of disorder and the resulting random tunneling.

The final outcome of the Kane-Fisher-Polchinski theory [49, 50] is a picture of decoupled charge and neutral modes with the action

$$\frac{S}{\hbar} = -\int dxdt \left\{ \frac{1}{4\pi}[-\partial_t\varphi_n\partial_x\varphi_n + v_n(\partial_x\varphi_n)^2] + \frac{3}{8\pi}[\partial_t\varphi_c\partial_x\varphi_c + v_c(\partial_x\varphi_c)^2] \right\} \quad (8)$$

where the charge mode $\varphi_c$ carries charge and energy *downstream*, and the *neutral mode* $\varphi_n$ carriers energy without net charge in the opposite, *upstream*, direction. The coefficient $\frac{3}{8\pi}$ in the action of the charge mode encodes the correct electrical conductance $2e^2/3h$. All the allowed perturbations that can be added to this action, such as random interaction of the charge and neutral modes, are irrelevant in the renormalization group sense.

Kane and Fisher extended the same picture to other states with *upstream* modes (hole-conjugate states); namely, states at the filling factors $\frac{p+1}{2p+1}$ [49]. In particular, they discovered that these states can be described in terms of a single *downstream* charged channel and *p* decoupled *upstream* neutral modes.

What are the experimental consequences of the *upstream* neutral modes? If the edge is longer than the electrochemical equilibration length between the counter-propagating channels, electrical current can only flow *downstream*, that is, in the direction of the charge mode. Energy, though, can go in both directions. Unfortunately, probing the energy flow is more difficult than probing charge currents. It was first proposed theoretically [51], and subsequently observed in experiments, that energy currents can be probed from excess charge noise with a zero net current [52]. We provide first a brief 'theoretical realization' of the two-QPC configuration [51] (Fig. 7), which allows an analytic calculation of the excess charge noise that results from *upstream* neutral modes.

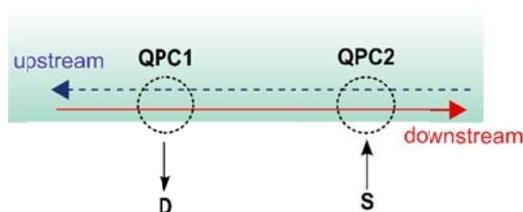

Fig. 7. A theoretical proposal to excite and measure an upstream neutral mode. The excitation of the mode takes place in **S** *downstream* QPC2. The energy propagates *upstream* (broken line) towards QPC1, where it fragments into (or heats up) the *downstream* charge mode (full line), with measured shot noise like fluctuations in **D**.

Imagine that charge can tunnel into and from the edge through QPC1 and QPC2 as shown in Fig. 7. Contact QPC1 is *upstream* relative to QPC2, with the charge mode running from QPC1 to QPC2 along the QHE edge. In the absence of *upstream* neutral modes (dotted line), all edge channels run in the *downstream* direction, while any *upstream* mode runs from QPC2 to QPC1. Here, QPC2 is biased while QPC1 is connected to unbiased drain D. If all edge modes move *downstream*, no energy and no charge will reach QPC1. For an edge supporting *upstream* neutral modes, the energy injected into the QHE edge at QPC2 travels in both directions along the edge, and thus arrives at QPC1. The energy arriving at QPC1 from QPC2 gives rise to excess noise in drain D. One can think of that noise as resulting from fragmentation of neutral particles to electron-hole pairs [52], or as an additional J-N noise due to the heating of QPC1 by the energy current [53]. Such excess noise is a smoking-gun evidence of an *upstream* neutral mode. While the exact setup of Fig. 7 allows analytic calculation of the excess noise [51], somewhat different setups have been employed to actually detect the *upstream* neutral modes, as discussed below [52-57].

### 4.2. *Experiments*

Proposals of how to detect the neutral modes involve measuring tunneling exponents in constrictions [49]; measuring thermal transport [58]; looking for resonances in a long constriction [59]; or looking for heating effects on shot noise [51, 60].

The first established experimental method involved a QPC constriction located *upstream* from an energized ohmic contact. An excited neutral mode, emanating from the '*hot spot*' at the source [61], propagates *upstream* along the edge and impinges on a partly pinched QPC. This leads to observed current fluctuations (Fig. 8) [52].

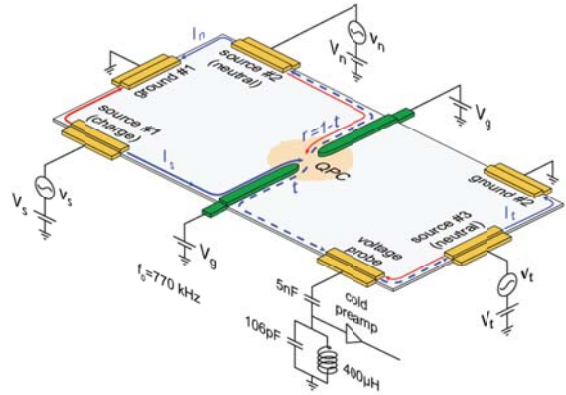

Fig. 8. An actual realization of the first experiment designed to measure the upstream neutral modes. Source #2, charged with $V_n$, injects charge current *downstream* (blue line, $I_n$), and a neutral mode, excited by the 'hot spot', *upstream* (red line). It impinges on the QPC, and generates charge fluctuations moving *downstream* (broken blue line). Noise is measured at the amplifier. Other contacts allow different measurements (from Ref. [52]).

Similar data was found in all hole-conjugate states and also in the $v$=5/2 state [55]. The noise had a $t(1-t)$ dependence, with $t$ the transmission of the QPC (Fig. 9). This dependence was attributed to the fragmentation of the dipole–like neutral carriers that carried the *upstream* heat. One cannot rule out thermal noise that resulted from heating up the constriction [53].

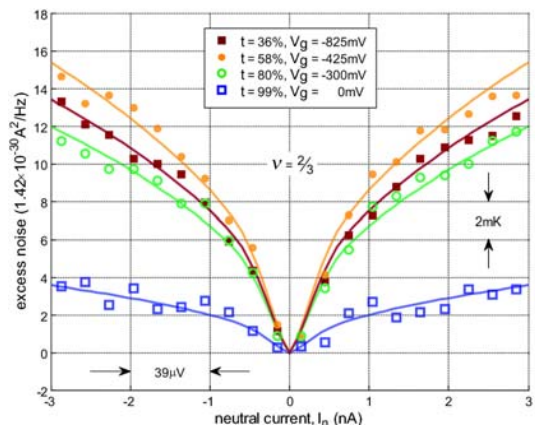

Fig. 9. Detection of an upstream neutral mode at filling $\nu = 2/3$. Shown is the excess noise, measured by the amplifier as a function of the current $I_n$ and the transmission $t$ of the QPC (see Fig. 8). The noise is proportional to $I_n$ and approximately to $t(1-t)$ (from Ref. [52]).

With a *downstream* charge mode and an *upstream* neutral mode impinging simultaneously on the QPC, the shot noise exhibited a higher particle temperature and a smaller Fano factor [52].

While a direct impingement of an *upstream* neutral mode on a large drain contact was not expected to lead to charge fluctuations, they were found in a subsequent experiment [10]. Simple heating cannot account for this noise, as the contact is macroscopic and, thus, hardly should heat up. Yet, this phenomenon can be understood if one takes into account the cold *downstream* channel that counter-propagates with respect to the *upstream* neutral mode. Inter-mode particle-hole excitations are created along the path of the neutral mode, and close to the hot-spot of the source contact, the charge excitations will be reabsorbed by the source. However, close enough to the drain, the fluctuations will be absorbed by the drain [62].

Since neutral modes carry heat, the most direct way to detect them is to measure the temperature increase at the measurement point *upstream* from the heat source. This can be done, for example, with quantum dot (QD) thermometry, that is, by allowing a neutral mode impinge on a QD and measuring the 'Coulomb blockaded' conductance peaks [54, 56]. Moreover, employing QDs, one can convert the temperature difference between the input and output QPCs into charge current [56].

It is worth emphasizing the fact that neutral modes were found along short edges in many particle-like states (such as $\nu$=1/3, 2/5, 1), which are not expected to support topologically protected upstream modes [10, 63]. Such upstream modes are not topological, and emerge due to edge-reconstruction, which, in turn, depends on the softness of the edge potential [9, 15]. Efforts to suppress these modes by sharpening the edge potential failed so far [63].

## 5. Interference of edge modes

The most direct way to ascertain the statistics of a fractional state is to braid its quasiparticles via interference experiments. Exploiting the chiral motion of the electrons in edge channels, an electronic analogue of the ubiquitous optical Mach-Zehnder interferometer (MZI) [64, 65] and an analogue of the Fabry-Perot interferometer (FPI) [66-69] were constructed (Fig. 10). While the MZI is a two-path interferometer, the FPI interferes multiple circulating paths. The two geometries provide complimentary experimental approaches. An FPI is simpler, but an MZI is more robust to effects of the bulk between the edges of the interferometer, as discussed in Sections 5 and 6.

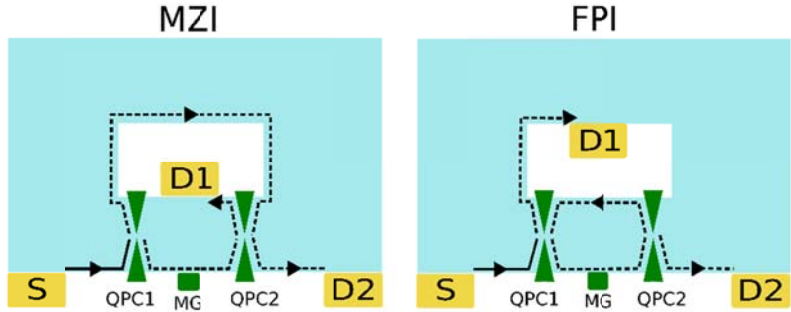

Fig. 10. Schematic realizations of a Mach-Zehnder Interferometer (MZI) and a Fabry-Perot Interferometer (FPI). Blue regions are the 2DEG, green regions are the QPCs, and white regions are etched (*vacuum*). Note different topology due to the location of drain D1.

Interferometry in FQHE appears to be challenging. Some of the results, observed mostly in the FPI, pose difficulties in their interpretation. In fact, even the interpretation of interference in the IQHE proved challenging. In this chapter, we focus on the IQHE. Two puzzling regimes will be highlighted later.

**5.1.** *Mach-Zehnder Interferometry*

Schematics of the electronic MZI is depicted in Fig. 10 [64]. Two QPC constrictions split and recombine the impinging edge channel, and two *ohmic contacts*, **D1** and **D2**, serve as drains. In an actual realization, with its schematics shown in Fig. 11, the inner contact, **D2** is connected to the outside circuit via an 'air bridge'. A phase difference $\varphi$ between the two paths is introduced via the Aharonov-Bohm (AB) effect, $\varphi_{AB} = 2\pi AB/\Phi_0$, where $A$ is the area enclosed by the two paths. A *modulation gate*, **MG**, is used to tune the area $A$.

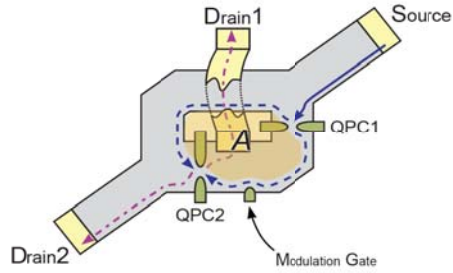

Fig. 11. A typical structure of an MZI. Note the inner contact (D1, in the orange region of the 2DEG), connected with an air bridge to an outside ground. A *modulation gate* is used to change the area enclosed by the two interfering trajectories (dotted lines).

We review briefly the operation of the interferometer. The conductance, from source to drain, is determined by the corresponding transmission probability $T_{SD}$, which, in turn, depends on the interference between the two paths. When the system is tuned to filling $\nu=1$, a single edge channel carries the current. Neglecting decoherence processes, with transmission (reflection) amplitude $t_i$ ($r_i$) of the $i^{th}$ QPC fulfilling $|r_i|^2+|t_i|^2=1$, the collected currents at **D1** and **D2** are, respectively,

$$I_1 \propto T_{SD1} = t_1^2 t_2^2 + r_1^2 r_2^2 + 2|t_1 t_2 r_1 r_2|\cos(\varphi_{AB}+\gamma) \qquad (9)$$

with $I_2$ the complementary current, $T_{SD1}+T_{SD2}=1$, and the phase $\gamma$ independent of the magnetic field. The visibility is defined as: visibility=$(I_{max}-I_{min})/(I_{max}+I_{min})$, and for $|t_2|^2=0.5$, visibility=$2|t_1|(1-|t_1|^2)^{1/2}$. We address below a few highlights, which reveal unexpected behavior in the integer regime.

### 5.1.1. *Lobe structure*

The highest visibility obtained in an MZI was at bulk filling $1<\nu_B<2$. In the non-linear regime, with a DC voltage added to a small AC signal, the visibility evolves in an unexpected fashion. Figure 12 shows the visibility and the phase of the AB oscillations at $\nu_B=2$ as a function of the applied DC voltage. The visibility oscillates between a maximum value and zero - in a lobe fashion [64]. Three striking features are observed: ***i***. The lobes scale in the source DC voltage is $V_S\sim15\mu V$ (diminishing with lowering the magnetic field). The scale of the overall decaying envelope is $V_S\sim30\text{-}50\mu V$. These energy scales are far smaller than any other energy in the IQHE regime. Moreover, diluting the impinging current on the MZI (with an external QPC) quenches the lobes pattern; ***ii***. The phase of the AB oscillations is rigid in each lobe, being independent of the energy (DC voltage). Crossing the zero visibility, at a certain value of $V_S$, the phase flips by π. ***iii***. The visibility shows exponential decay with increasing temperature and length of the MZI arms [65, 70-72]. These results, which cannot be explained by a single-particle model, are worth a further consideration.

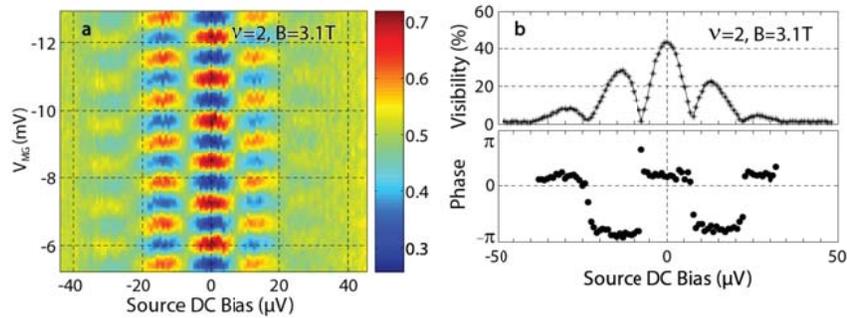

Fig. 12. Aharonov-Bohm (AB) interference in an MZI. (a) Oscillations of the conductance as a function of the area (tuned by the modulation gate, $V_{MG}$) and an added DC source bias (arbitrary scale: red – high visibility, blue – low visibility). (b) The visibility of the AB oscillations as a function of the DC bias. Note the *lobe structure*, with a constant phase of the oscillations throughout each lobe and with an abrupt reversal at the nodes (from Ref. [65]).

Naively, when treating the interfering integer edge channel as a non-interacting Fermi gas, one would expect a rather weak dependence of the visibility on the source voltage. However, inter-channel interaction, and charge fluctuations (due to partitioning the incoming current with the QPCs) may complicate the simple non-interacting picture [73-77].

Two leading theoretical models catch some of the features of the lobes structure. One considers short-range inter-channel interaction at filling factor two, as well as long-range intra-edge interactions. Inter-channel interaction gives birth to two modes, slow and fast. Roughly speaking [78], the scale of the lobes is comparable to $eV = \hbar v/L$, with $v$ the velocity of the slow mode. The overall envelope is related to the fluctuations in the number of electrons in each arm of the MZI [79, 80]. However, the model fails to explain the phase rigidity away from $L_1 = L_2$, and especially at filling one.

Another approach assumes ad-hoc bunching of two electrons with interference of all possible trajectories [81]. While this assumption seems unjustified, such bunching was found in experiments with FPI (see Section 5.2.3).

### 5.1.2. *Phase averaging versus decoherence*

Interference can be quenched in two different ways: First, *decoherence*, e.g., spin flips in one of the two paths. Second, *phase averaging*: particles interfere but each with a different phase (due to fluctuations of the phase in time or phase dependence on the energy). These two mechanisms can be distinguished by measuring the shot noise of the interfering signal. Assume a phenomenological parameter that accounts for decoherence, $k$, with $t_1^2=0.5$; then, $I_1 \propto T_{SD1}=0.5+k[t_2^2(1-t_2^2)]^{1/2}\langle\cos\varphi_{AB}\rangle$, where the angle brackets denote phase averaging. Obviously, either when $k=0$ (total decoherence) or when $\varphi_{AB}$ spans a $2\pi$ range, the interference term (second term) is expected to vanish. However, the shot noise $S_1 \propto T_{SD1}(1-T_{SD1})=1/4-k^2 t_2^2(1-t_2^2)\langle\cos^2\varphi_{AB}\rangle$, will have the following dependence on $t_2$: *i.* For $k=0$, $S_1$ is independent of $t_2$; *ii.* For $k>0$, the $\langle\cos^2\varphi_{AB}\rangle$ term will not vanish upon averaging, leading to $S_1=1/4-k^2 t_2^2(1-t_2^2)/2$. These two extreme behaviors may help in distinguishing between the dephasing mechanisms. Thus far, only $k>0$ was found (e.g., [64]).

### 5.1.3. *Melting of the interference*

An observation of interference of anyons would be a step towards understanding their anyonic statistics. However, the complex edge structures of FQHE states, hosting counter-propagating charge and neutral channels, have been suspected to prevent the observation of the much sought after interference of anyons. Specifically, the interference of the outer-most edge channel in the MZI was found to diminish gradually as the filling was lowered towards $v_B$~1, followed by an apparent increase of the excitation of a 'non-topological' *upstream* neutral mode [63, 81]. The simultaneous emergence of a neutral mode and a $\nu_{QPC}=1/3$ conductance plateau in a partly pinched QPC suggested a reconstruction of the edge mode into two modes, $\nu = 2/3$ and $\nu = 1/3$, which leads to the formation of an upstream neutral mode. The $\nu_{QPC}=1/3$ conductance plateau was found to persist for the range of bulk fillings ½<$\nu_B$<1.5. Uncharacteristically, shot noise was measured on the conductance plateau, resulting from the partitioned *upstream* neutral mode [16, 63]. Moreover, and even more surprisingly, particle-like states in the range of $\nu_B$<1/2 were also accompanied by *upstream* neutral modes - independently of efforts to sharpen the edge-potential [10, 63].

### 5.2. *Fabry-Perot interferometry*

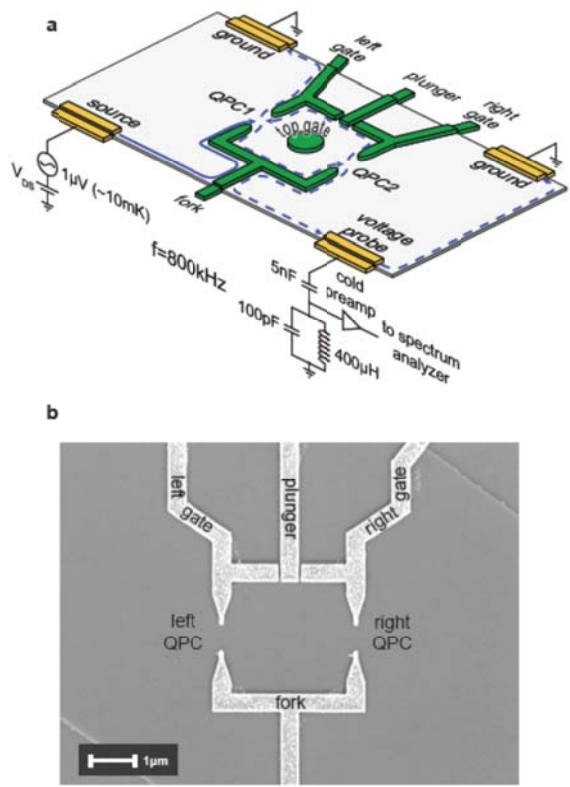

Fig. 13. A typical FPI. (a) A two-QPC schematic configuration of the FPI, with a single source and two drains (one at the amplifier and one at the ground). The broken lines represent the partitioned channel. Multiple trajectories circulate around the interior of the FPI (not shown). A plunger gate is used to change the enclosed area of the trajectories. A top gate is sometimes added for screening and testing. (b) An SEM micrograph of a typical FPI. The light gray areas are the gates forming the QPCs and the plunger gate (from Ref. [69]).

An FPI is a large quantum dot (QD). Like an MZI, it is formed by two QPCs but without an internal drain contact. The transmitted and reflected currents are absorbed by two remote detectors. Interference takes place among multiple trajectories that circulate around the inner perimeter of the FPI (Figs. 10 and 13).

In the lowest order, when the backscattering in both QPCs is weak, the interference takes place only between two dominant partial waves, with $I_1$ the transmitted current,

$$I_1 \sim T_{SD1} \approx t_1^2 t_2^2 + 2t_1 t_2 r_1 r_2 cos(\varphi_{AB} + \gamma) \qquad (10)$$

where the parameters are defined above ($I_2$ is the complimentary reflected current).

Like in the MZI, the relative phase between consecutive rounds is controlled by varying the enclosed AB flux (with the magnetic field or a *modulation gate*). When Coulomb interaction is dominant, the enclosed charge tends to be constant, maintaining charge neutrality. This regime is called the *Coulomb Dominated* (CD) regime [67-69, 82, 83].

### 5.2.1. *Aharonov-Bohm and Coulomb-dominated interference*

In the purely AB regime Coulomb interactions are negligible, and the 'flux containing area' remains nearly constant with changing the magnetic field. Since the degeneracy of the LLs increases, charge is added to the liquid and the periodicity is expected to be that of a flux quantum. Here, since screening is highly effective, we assume that no quasiparticles are added to the bulk. When interactions dominate, the behavior of the FPI in the CD regime is quite different. For a general filling factor, there are fully reflected, partially transmitted, and fully transmitted edge channels in the QPCs that form the FPI. The number of fully reflected channels is the same as the number of the circulating channels within the interior of the FPI. Numbering the fully reflected channels by $f_R$ and the fully transmitted ones by $f_T$, the total number of participating channels is $f_{edge}=f_T+1+f_R$ - being determined by the bulk filling factor $\nu$ (for IQHE $f_{edge}=\nu$). Each configuration can be denoted as ($\nu, f_T$). The experiments [69] revealed the following:
1. Each of the edge channels could be made to interfere independently of the others;

2. The periodicities in the magnetic field $\Delta B$ and the modulation-gate voltage $\Delta V_{MG}$ do not directly depend on the magnetic field (or the filling factor) and the transmission amplitudes $t_1$ and $t_2$, but only on $A$ and $f_T$:

$$\Delta B(B, V_{MG}, f_T, t_1, t_2) = \Delta B(f_T) = -\frac{\phi_0}{A \cdot f_T}, \quad (11)$$

as was experimentally found to hold for a wide range of QPC's transmission. Similarly, the periodicity in the modulation gate voltage depends only on $f_T$:

$$\Delta V_{MG}(B, V_{MG}, f_T, t_1, t_2) = \Delta V_{MG}(f_T) \quad (12)$$

with $\Delta V_{MG}$ monotonically increasing with increasing $f_T$ (see an exception in [69]);

3. For $f_T=0$, the enclosed flux is independent of the magnetic field, indicating that the area of the FPI is proportional to $1/B$ (Fig. 14);
4. For $f_T>0$, the enclosed flux decreases with increased magnetic field, indicating that the area of the FPI shrinks faster than $1/B$.

In Appendix C a more detailed elaboration of this behavior is presented.

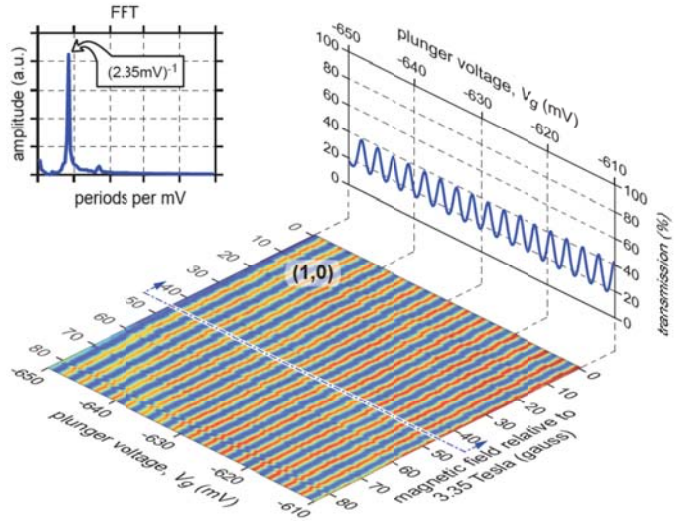

Fig. 14. Interference of the edge channel at filling $\nu=1$ in the FPI shown in Fig. 13. The interference is dominated by Coulomb interactions. While changing the area results in nice oscillations (with a periodicity corresponding to adding/removing an electron), no magnetic field dependence is observed (from Ref. [69]).

5.2.2. *The AB – CD mixed regime*

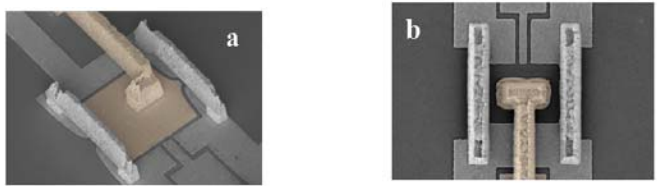

Fig. 15. Two realizations of screened FPIs. (a) A grounded top gate screens the interior of the FPI, allowing adding or subtracting charge with little energy cost. (b) A grounded ohmic contact in the center of the FPI eliminates the restrictions forced by Coulomb interactions. Charge can be added or drained from the bulk with no energy cost. While the configuration in (b) always exhibits AB interference, configuration (a) exhibits AB interference only above a critical area (from Ref. [84]).

Several ways to overcome the CD behavior were suggested: 1) Covering the sample with a top gate in order to increase the total capacitance (Fig. 15a) [68, 69, 85], or adding an internal screening layer [86]. 2) Placing an ohmic contact in the bulk to break its isolation and allow carrier drainage (Fig. 15b) [84, 87]. This method was found to be more effective than placing a top gate. 3). An intermediate coupling regime was achieved by placing a large ohmic contact outside the perimeter of the interferometer [88]. A checker-board pattern of the conductance (as a function of $B$ and $V_{MG}$), with periodicities that correspond to both AB and CD regimes, is typically observed (Fig. 16) [88]. This configuration is promising, since it allows for construction of small interferometers; yet, provides information on the coherent component of the interference pattern. Note, the physics behind the partial suppression of the Coulomb interaction is not yet understood.

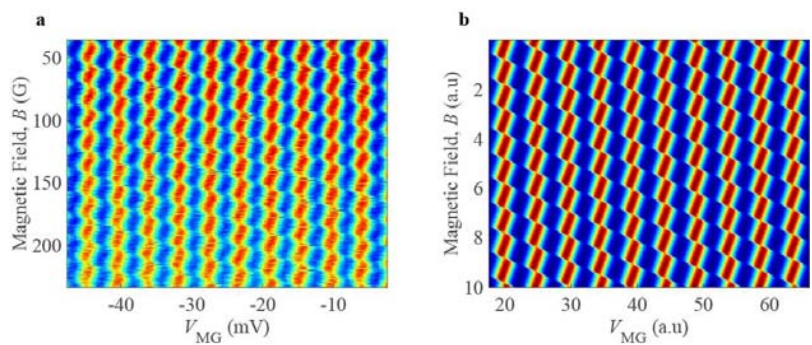

Fig. 16. The mixed AB-CD regime. (a) Measured checker-board pattern of the interference. (b) Calculated checkered-board (from Ref. [88]).

### 5.2.3. *The Pairing effect*

Electron 'pairing' is a rare phenomenon appearing only in a few unique physical systems, e.g., superconductors and Kondo-correlated quantum dots [89, 90]. Yet, in an FPI in the AB regime (with suppressed Coulomb interactions), an unexpected, but robust, electron 'pairing' was found in the IQHE regime. The pairing took place within the outermost interfering edge channel in a wide range of bulk filling factors, $2<\nu<5$ (Fig. 17) [84, 86, 87]. The main observations were: *i.* High visibility AB conductance oscillations with the magnetic flux periodicity of $\Phi_0/2=h/2e$ (instead of the ubiquitous $h/e$ periodicity); *ii.* The interfering quasiparticle charge was $e^*\sim2e$, as determined by shot noise measurements; *iii.* The $h/2e$ periodicity was fully dephased when the first inner (adjacent to the outermost) edge channel was dephased (while keeping the interfering edge channel intact). This was a clear observation of inter-channel entanglement; and *iv.* The AB flux periodicity was exclusively determined by the enclosed area of the first inner (adjacent to the outermost) channel and *not* by the outermost interfering channel.

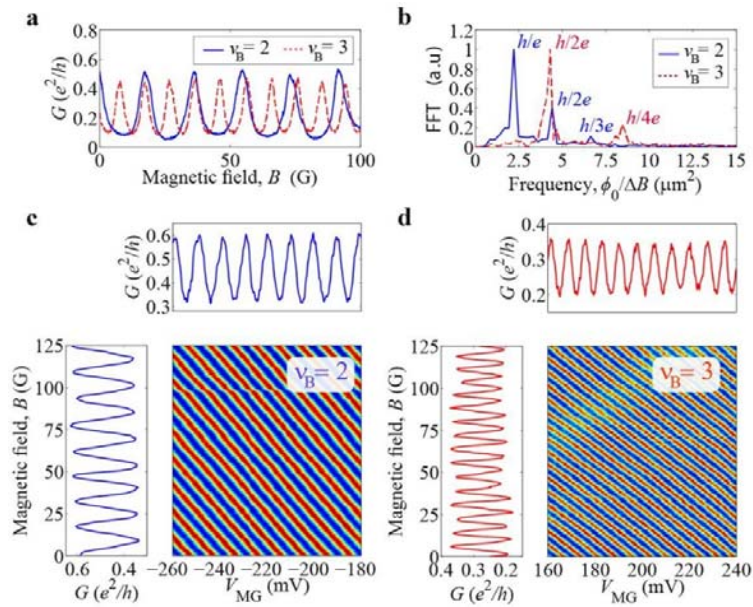

Fig. 17. The *pairing* phenomenon detected in the screened FPI shown in Fig. 15a. (a) Interfernece oscillation of the outer-most edge channel at $v$=2 and $v$=3. (b) FFT of the oscillations shows an exact factor of two in the periodicities. (c) and (d) show the 2D plot of the conductance as a function of $B$ and $V_{MG}$, with constant phase lines proving pure AB interference (from Ref. [87]).

An extensive set of measurements suggests the formation of a neutral edge mode due to strong interaction between the two outer edge channels. The neutral mode plays a crucial role in the pairing phenomenon; however, the exact mechanism for this phenomenon is not clear. The pairing may not be a curious isolated phenomenon, but one of many manifestations of unexpected edge physics in the quantum Hall regime (see the above lobe-structure).

## 6. Non-abelian order of the $v = 5/2$ state

### 6.1. *Fractional statistics*

While the fractional Hall conductance goes hand-in-hand with the fractional filling, the fractional charge goes hand-in-hand with *fractional statistics*, which is neither Bose nor Fermi. We illustrate this point with an example of the Laughlin state at $v = 1/3$ in an MZI. The elementary excitations carry the charge $e/3$ and the edge consists of a single chiral mode. Imagine that the FQHE liquid fills the blue region of an MZI (Fig. 18 and Section 5). A gedanken experiment with the MZI can demonstrate that the $e/3$ quasiparticles cannot be fermions or bosons.

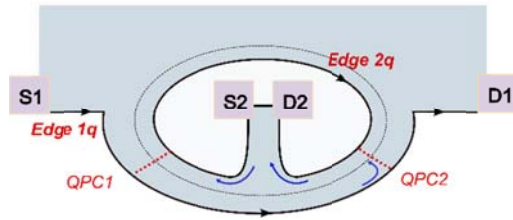

Fig. 18. A simplified schematic picture of a Mach-Zehnder interferometer. Quasiparticles tunnel between the inner and outer edges at QPC1 and QPC2. Drains D1 and D2 absorb the interfering quasiparticles.

Recall Eq. (9), with the phase-dependent term in the transmission, $t_{SD1} = \eta \cos(\varphi_{AB}+\gamma)$, with $\varphi_{AB} = 2\pi\phi/\phi_{1/3}$ and $\phi_{1/3} = 3\Phi_0$. The transmission probability, and hence, apparently, the current is periodic in the magnetic flux $\phi$ with the period of three flux quanta. This contradicts the rigorous Byers-Yang theorem [91], which states that the period cannot exceed $\Phi_0$. The theorem expresses the fact that an FQHE liquid is made of electrons, and any integer number of flux quanta is invisible to electrons that make a circle around the enclosed flux.

In order to solve the paradox one must recall the notion of fractional statistics. Specifically, the quasiparticle wave function accumulates a non-trivial phase when encircling another quasiparticle. Then the statistical phase $\gamma$ in Eq. (9) in combination with $\varphi_{AB}$ leads to a total phase that is proportional to the number $n$ of $e/3$ quasiparticles inside the dashed circle (Fig. 18). Each tunneling event changes $n$, since the tunneling quasiparticles are absorbed by drain D2 inside the dashed loop. Hence, the oscillating transmission depends on $t_{\text{SD2}} = \eta \cos\left(\varphi_{AB} + \frac{2\pi n}{3}\right)$ returning to its initial value after three consecutive tunneling events of quasiparticles. The three events transfer to D2 the combined charge of one electron. This is crucial for the "single-electron" periodicity, demanded by the Byers-Yang theorem. The drain current can be found as the ratio of that charge and the average time it takes for three tunneling events, $I_{\text{SD2}} = \frac{e}{\frac{1}{p(1)} + \frac{1}{p(2)} + \frac{1}{p(3)}}$, where $p(i)$ is the transmission probability after $i$ consecutive tunneling events (modulo $i=3$). This result [92] is consistent with the Byers-Yang theorem [91].

### 6.2. *Non-Abelian statistics and the 5/2 liquid*

In the previous Section we considered an example of *abelian statistics*: particles accumulate non-trivial phases when they go around each other, but combining (or fusing in the usual terminology) any two anyons results in a unique particle type, as two fermions always fuse into a boson. However, *non-abelian anyons* have multiple fusion channels [93, 94]. Two identical anyons can form multiple bound states, which can be understood as composite particles - and different composite particles can exhibit at least two different statistics. Such property is seen as valuable for quantum computing [94, 95]. The existence of non-abelian anyons was first conjectured at the FQHE filling of $\nu=5/2$, consequently, attracting intense interest from experimentalists in recent years.

In studying this fractional state, geometric resonance measurements observed cyclotron motion of CFs [96, 97], suggesting that CFs are present at $\nu = 5/2$. This leads to a theoretical challenge. Indeed, in a mean-field description, CFs of even-denominator states move in an effective zero magnetic field. This seems to conflict with the observed energy gaps at $\nu = 5/2$ and $\nu = 7/2$, since the liquids of weakly interacting fermions in a zero field are gapless, as is indeed the case at $\nu = 1/2$ and $\nu = 3/2$. To explain the observed gapped states, one has to assume that CFs form Cooper pairs [98].

The language of Cooper pairs allows a prediction of the quasiparticle's charge. In a rather simplistic picture, the Cooper pairs are interpreted as independent bosonic particles of charge 2*e*, while the two filled integer Landau levels are ignored. Taking $\nu = 1/2$ for electrons translates into the filling factor $\nu' = 1/8$ for the charge-2*e* bosons. Just as the quasiparticle charge is $\nu e$ in the Laughlin state of the filling factor $\nu$ for electrons, now the quasiparticles are of charge $2e\nu' = e/4$. In other words, the paired state allows insertion of a half-vortex with the corresponding charge *e*/4. This quasiparticle charge is consistent with several experiments [99-102].

What is the exchange statistics of the *e*/4 anyons? Following Laughlin's argument (Section 3), the insertion of one flux quantum generates an excitation of charge *e*/2 at filling *ν*=1/2. Thus, one might naively think that one just needs to insert two flux quanta to get an electron with charge *e*. However, the argument for the exchange phase here is different. The exchange of electrons generates an exchange phase of $\pi$; achieved also by moving one electron along half a circle with the center at the other electron. Yet, such half-encircling of the two inserted flux quanta in the center leads to an accumulated Aharonov-Bohm phase of $2\pi$ – suggesting that flux insertion generates a charge-*e* boson (instead of an electron). Hence, a combination of a hole of charge −*e* with such boson yields a *neutral fermion*. In the composite fermion language, this is just an unpaired composite fermion. Such fermions can be combined with charge-*e*/2 excitations, which thus come in two species. In other words, we can build two combinations of two *e*/4 anyons. Two logical possibilities are now open: 1) The system has more than one sort of *e*/4 particles; 2) only one sort of *e*/4 anyons exists. In the first case, two types of *e*/2 anyons could be built from different combinations of elementary anyons. In the second case, there must be two ways to fuse two identical particles. This is a signature of non-Abelian statistics.

To refine the list of possibilities, we turn to the edge physics. Naively, the edge action is very similar to the actions at $\nu = 1$ and at $\nu = 1/3$.

$$\frac{S}{\hbar} = -\frac{2}{4\pi} \int dx\, dt [\partial_t \varphi \partial_x \varphi + v(\partial_x \varphi)^2] \,, \tag{13}$$

with the coefficient of 2 reflecting the electrical conductance $e^2/2h$ of the half-filled Landau level. To identify the electron operator at the edge, an operator that destroys charge *e* must be constructed. This follows the logic of Appendix A and yields the operator $\exp(2i\varphi(x))$. Alas, this answer cannot possibly be correct, as it satisfies bosonic commutation relations instead of the expected fermionic ones. Hence, we have to assume that the edge has an additional neutral fermionic mode (or modes, $\psi_k$),

$$\frac{S}{\hbar} = -\frac{2}{4\pi}\int dx\,dt[\partial_t\varphi\partial_x\varphi + v(\partial_x\varphi)^2] + \int dxdt \sum i\psi_k(\partial_t + v_k\partial_x)\psi_k \,, \quad (14)$$

with electron operators constructed as $\psi_k \exp(2i\varphi(x))$. The operators $\psi_k$ do not have to be complex Dirac fermions. They can also be real *Majorana fermions*, which are real or imaginary parts of Dirac fermions. Indeed, $\psi_k$ account for the fermionic parity of the system, while the total number of fermions is counted by the Bose field $\varphi(x)$. Moreover, since a Dirac fermion is just a combination of two Majorana fermions, then, without loss of generality, $\psi_k$ can represent Majorana operators.

Opposite signs of $v_k$ correspond to opposite chiralities of edge modes. With $v_k$ having opposite signs, pairs of counter-propagating Majorana modes would be localized (gapped) by disorder. Thus, we can assume the same sign for all $v_k$. Different allowed edge structures are distinguished by the Chern number $v_C$, that is, by the net number of the Majorana edge modes [103]. $v_C$ is negative if the propagation direction of the Majorana modes is *upstream*, while the central charge, proportional to the fully equilibrated thermal conductance, is c=1 + $v_C/2$.

According to Kitaev's classification [95], all topological orders can be distinguished by the Chern number. In particular, an *even* $v_C$ corresponds to abelian statistics while an *odd* $v_C$ to a non-abelian topological order. To see the origin of this rule, the principle of *bulk-edge correspondence* is discussed, though heuristically, in Appendix B. A rigorous approach is addressed in detail in Refs. [95, 103].

### 6.3. *Proposed 5/2 states*

Not all choices of the Chern number are equally interesting or likely to occur in nature. All the realistic possibilities in the literature have a small $v_C$. Moreover, according to Kitaev's 16-fold way [95], fractional statistics depends only on $v_C$ mod 16. Table I lists the states that received most attention.

Table I. Proposed states with low Chern numbers.

| $v_C$ | -3 | -2 | -1 | 0 | 1 | 2 | 3 |
|---|---|---|---|---|---|---|---|
| order | anti-Pfaffian | 113 | PH-Pfaffian | $K=8$ | Pfaffian | **331** | $SU(2)_2$ |
| Refs. | [58, 104] | [105] | [104, 106-109] | [110] | [93] | [111] | [112, 113] |

The non-abelian *PH-Pfaffian* order is particularly interesting due to its *particle-hole symmetry* [106]. The filling factor of 1/2 can be interpreted in two ways: as a half-filled Landau level of electrons or a half-filled Landau level of holes. The particle-hole (*PH*) transformation changes electrons into holes and vice versa. In general, the *PH*-transformation changes the topological order since the topological orders for the electrons and holes do not have to be the same. We proceed in the spirit of Fig. 2, which illustrates the effect of the *PH*-transformation on the edge modes of the 2/3-liquid of electrons, which can be also thought of as Laughlin's *v*=1/3 state of holes. The figure shows how the *v*=2/3 edge mode hosts a *downstream* integer *v*=1 channel and an *upstream* *v*=1/3 channel. The same trick applies at $v = 5/2$, where the *downstream* integer channels are ignored. For example, start with the *Pfaffian* edge, which contains a *downstream* charged *bosonic* mode (with conductance $e^2/2h$) and one *downstream* Majorana mode. The *PH*-transformation yields a *downstream* integer channel (with conductance $e^2/h$), an *upstream* charged *bosonic* mode (with conductance $e^2/2h$), and an *upstream* Majorana mode. The same arguments as presented in Appendix A show that the two bosonic modes (a *downstream* integer and an *upstream* with conductance $e^2/2h$) reorganize into a *downstream* charge mode with conductance $e^2/2h$, and an *upstream* neutral mode (preserving the central charge). The latter can be fermionized and represented as two *upstream* Majorana modes. Accounting for one more *upstream* Majorana, we get $v_C = -3$, that is, the *anti-Pfaffian* state. The *PH-Pfaffian* state is unique in not changing its structure under the *PH* transformation. The edge contains a downstream charged mode of conductance $e^2/2h$ and an upstream Majorana fermion. The PH transformation results in a downstream mode of conductance $e^2/h$, upstream mode of conductance $e^2/2h$, and a downstream neutral Majorana mode. The two charge modes reorganize into a single downstream charged mode and an upstream neutral boson. The boson can be fermionized and represented as a combination of two upstream Majorana modes. Interaction with the downstream Majorana gaps one of them out. We are left with the initial edge structure of the *PH-Pfaffian* liquid.

Past experiments did not provide a way to determine the topological order of the *v*=5/2 state, and the focus had been on numerical calculations. Numerical evidence was produced for several orders [114-116], and eventually, the numerical debate concentrated on the Pfaffian ($v_C = 1$) and anti-Pfaffian ($v_C = -3$) states [116-118]. Yet, as shown below, experimental evidence has increasingly pointed towards the PH-Pfaffian order ($v_C = -1$) [107].

### 6.4. *Experiments at* $\nu = 5/2$

As discussed above (Section 6.2), multiple groups have obtained results compatible with the quasiparticle charge of *e*/4. This does not shed light on the topological order since the same charge is predicted for all orders of the 16-fold way. Other existing probes of bulk properties unfortunately shed little light on topological order. More useful information comes from edge probes [103]. The observation of *upstream* heat transport at $\nu = 5/2$ provides evidence of a negative Chern number and is inconsistent with numerically supported Pfaffian order. As discussed in Ref. [107], tunneling experiments appear inconsistent with the numerically supported anti-Pfaffian state [119-122]. On the other hand, those experiments seem compatible with the *PH-Pfaffian* order. A natural way to distinguish the *PH-Pfaffian* state from numerically supported candidates is based on thermal conductance measurements, discussed in the next Section. Another approach is interferometry.

The most obvious and deterministic way to measure the braiding statistics of quasiparticles involves interferometry. For non-abelian states, the FPI is expected to exhibit the *even-odd effect* [123, 124]. In other words, the contribution of the interference to the current, which follows $\sim\cos(\varphi + \gamma)$, can only be observed if the total number of *e*/4 quasiparticles inside the interferometer is even. This can be understood in the following way: Let the number of trapped anyons be odd, say one. The trapped anyon has two fusion channels with the anyon at the edge mode. The probabilities of the two fusion outcomes turn out to be the same, while the statistical phases differ by $\pi$. This leads to total destructive interference of the two fusion channels and no magnetic flux dependence of the current. The argument does not apply for systems with an even number of trapped anyons, where the interference should be restored because then there is only one fusion possibility. As an example, the existence of a single fusion channel is most obvious when the number of trapped anyons is zero since then the edge anyon can fuse with nothing in only one way. This beautiful effect is a striking signature of a non-abelian topological order; yet, it cannot distinguish among different non-abelian orders. Surprisingly, under certain conditions it might not be able to distinguish non-abelian orders from abelian ones, since the existence of two types of abelian *e*/4 particles can mimic two fusion channels of non-abelian anyons [125].

Interference experiments performed with a small FPI with the *ν*=5/2 state are consistent with non-abelian statistics [126]; yet, their theoretical interpretation has proved difficult [127]. The MZI [64] is free from some of the challenges of the FPI. Hence, it should offer unique signatures for various topological orders [92, 128-133]. The theory of the MZI as an anyonic interferometer is considerably more involved than that in the FPI case [92]. Here, we only mention a particularly striking behavior, predicted for the *PH-Pfaffian* state [107]: the transmitted current through the MZI is not expected to depend on the magnetic field, while the noise of that current diverges at some values of the field. The very different behavior of the current and noise in the MZI reflects a long-term memory effect (Section 6.1), absent in the FPI.

## 7. Thermal transport

### 7.1. *Theoretical arguments*

It has long been known that in the QHE both electrical and thermal conductance are quantized [134-136]. Yet, the first observation of quantized heat current in the QHE regime came more than three decades after the discovery of quantized electric transport reflecting the difficulty of probing neutral modes.

The fundamental limit of quantum heat flow was discovered around the same time as FQHE [134-136]. The simplest example of the quantization is a 1D quantum wire. Let $v(k) = \frac{d\varepsilon(k)}{\hbar dk}$ be the velocity of non-interacting fermions in a 1D conductor. Assume that they enter a 1D ballistic quantum wire from the left at a temperature $T_L$ and from the right, where the fermions are at a temperature $T_R$. The heat flow through the wire is

$$J_T = \int \frac{dk}{2\pi} \varepsilon(k) v(k) [n(T_L) - n(T_R)] = \frac{\pi^2 k_B^2 (T_L^2 - T_R^2)}{6h} \quad (15)$$

with *n*(*T*) the Fermi distribution at the temperature *T*. This yields the quantized thermal conductance $\kappa_0 = \frac{dJ_T}{d\Delta T} = \frac{\pi^2 k_B^2 T}{3h}$, where $\Delta T = T_L - T_R \to 0$ (linear regime). In contrast to the quantization of the electrical conductance in quantum wires, the result in Eq. (15) does not hold in the presence of inter-mode interaction among counter-propagating channels. Evidently, inter-channel equilibration will backscatter heat via the *upstream* channels. At the same time, quantization of the heat conductance in the QHE regime is still expected as long as the propagation length is substantially longer than the thermal equilibration length among all channels. Moreover, no heat exchange between the two spatially separated edges of the sample should be allowed. Note that the same quantization of heat transport as in IQHE channels holds for abelian FQHE channels [135, 137]. The reason is simple: the action given in Eq. (4) for an FQHE channel can be reduced to the action of IQHE in Eq. (3) by a rescaling of the Bose field $\varphi$. The same argument would not work for charge transport since the rescaling would affect the definition of the charge density. No such problem is encountered for the energy density since it is nothing else but the Hamiltonian density.

For the case of several non-interacting co-propagating chiral *downstream* edge channels, their thermal conductances simply add up:

$$\kappa = n\kappa_0 \qquad (16)$$

where *n* is the total number of the channels. Essentially, the same argument as for charge transport in Section 4 shows that interaction between co-propagating modes cannot change Eq. (16). In the presence of counter-propagating channels for a long distance, *n* becomes the difference between the numbers of the *downstream* and *upstream* channels (that is, the net *downstream* number of channels).

While the heat conductance does not distinguish states in the IQHE from abelian states in the FQHE, it distinguishes non-abelian orders from abelian ones. Indeed, for non-abelian orders, *n* in Eq. (16) is no longer an integer [136]. The example of Majorana fermions is most important in this context. A Majorana fermion is a real or imaginary part of a usual complex fermion, and hence, can be thought of as a half of a Dirac fermion. Thus, the thermal conductance of a Majorana channel is one-half of the thermal conductance of a Dirac channel, thus, contributing $\kappa_0/2$ to the thermal conductance. This suggests a tempting way to probe the topological order of the $\nu = 5/2$ state: all needed to know is the Chern number, and the thermal conductance is proportional to it.

### 7.2. *Experimental results*

A pioneering measurement of the thermal conductance in the IQHE regime was performed by Jezouin et. al. [138]. This work was extended to the FQHE regime by Banerjee et al. [139], proving the universality of the thermal conductance in 'particle-like' fractional states. Moreover, measurements were extended to the more intriguing 'hole-conjugate' states (i.e., $\nu$=2/3, 3/5, and 4/7), which support counter-propagating charge and neutral modes. Having this baseline, the developed methods allowed exploring the states in the second Landau level, and, in particular, the intricate $\nu$=5/2 state. The latter was found to support a fractional thermal conductance [140].

More specifically, the $\nu$=1/3 Laughlin state, with conductance $e^2/3h$ and quasiparticle charge $e^*$=$e$/3, was found to carry one unit of thermal conductance **1**$\kappa_0 T$, without a factor 1/3 [139]. The 'hole-like' states, $\nu$=3/5 and $\nu$=4/7, supporting a larger number of *upstream* chiral neutral modes (see Section 2.2), had a *negative* thermal conductance determined by the **net** chirality of all their edge modes (the *negative* sign of the heat flow was not measured) [135]. The most explored $v=2/3$ state is special. It has a *downstream* charge mode with conductance $2e^2/3h$ and an *upstream* neutral mode, so that the net number of the *downstream* modes is $n=0$. Since full equilibration takes place only at infinite propagation length (thermal transport is diffusive), the lowest thermal conductance was $\sim 0.25\kappa_0 T$ at some 40mK [139].

Similar measurements were performed in the second Landau level, testing the three major fractions, $v=7/3$, $5/2$, and $8/3$. While the odd denominator states were found to conduct heat as expected, the thermal conductance of the $v=5/2$ state was found to be **2.5**$\kappa_0$; suggesting the *PH-Pfaffian* order [140]. This order is not expected from numerical calculations (see discussion below).

### 7.3. *Experimental method*

The experimental setups adopted heating a small floating metallic contact, which in turn emitted edge modes towards a colder (grounded) contact [138-140]. Measuring the temperature of the floating contact was sufficient for the determination of the total thermal conductance.

A DC input, provided by a source current, heated the floating contact. The outgoing current was split into $N$ similar wide arms (mesas), with $n$ chiral channels propagating along the two edges of each arm. The dissipated power in the floating contact raised the contact's temperature in equilibrium to $T_m$, with the dissipated power being equal to the emitted power. The latter was carried away from the floating contact by phonons (to the bulk) and by the edge modes, $P_{diss}=\Delta P_{ph}+\Delta P_e$, respectively. Consequently, $P_{diss}=0.5 \cdot N \cdot n \cdot \kappa_o \cdot (T_m^2-T_0^2)+\beta\ (T_m^5-T_0^5)$, where $T_0$ is the electron temperature in the 'cold' contacts, and $\beta$ is the prefactor of the phonon emission (depends on the floating contact size) [141]. The phonon contribution became negligible in comparison with the electronic contribution for $T_m<30$mK and $T_0=10$mK (depending on $\beta$ [139]). At higher temperatures, the phonon contribution can be subtracted out by changing the participating number of the edge modes emanating from the floating contact. The contact temperature $T_m$ is deduced by measuring the *downstream* thermal noise (Fig. 19).

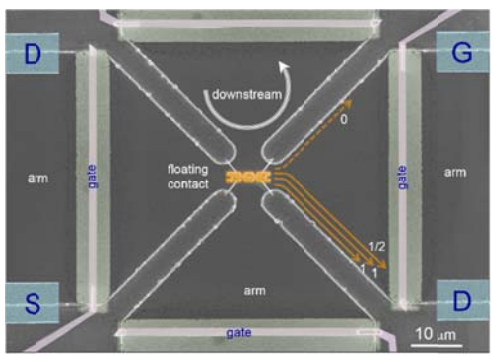

Fig. 19. The heart of the device used to measure thermal transport at filling *v*=5/2. The small ohmic contact (area 12μm$^2$) serves as the heated floating reservoir and injects currents into four arms. The effective propagation length (to a cold contact, not shown) in each arm is ~150μm. Arms can be pinched-off by negatively charging surface gates. As an example, the energy carrying edge modes that correspond to the *PH-Pfaffian* order are shown ('cold modes' are not shown). The solid orange arrows represent *downstream* charge modes; each carries $\kappa_0 T$ heat flux. The dashed orange arrow represents an *upstream* Majorana mode, carrying $0.5\kappa_0 T$ heat flux (from Ref.[140]).

### 7.4. *Experimental considerations*

Measuring the thermal conductance is more demanding and less accurate than measuring the electrical conductance. Here are a few points that highlight a few of the difficulties: ***i.*** Charge should fully equilibrate in the heated small floating contact; ***ii.*** The floating contact must be 'close to ideal'; namely, with a negligible reflection coefficient (to minimize producing excess shot noise); ***iii.*** The outgoing current must split equally between all the open arms; ***iv***. Bulk heat conductance must be negligible (in particular for the $v$ =5/2 samples, where the doping scheme can leave parallel heat conduction through the doping regions [140, 142, 143]); ***v***. The temperature must be stable during prolonged measurements (in particular at the lowest temperatures); ***vi***. The gain of the amplification chain must be carefully calibrated, as it is crucial in the determination of the temperature; ***vii.*** Energy loss to the environment (e.g., due to Coulomb interaction with metallic gates) must be minimized.

### 7.5. *Inter-channel equilibration*

Deviations from the expected thermal conductance were found at the filling factors *v*=2/3 [139] and *v*=8/3 [140]. At $v = 5/2$, the observed thermal conductance was ~$2.5\kappa_0 T$ at a range of temperatures, but increased to ~$2.75\kappa_0 T$ at a lower temperature. It is likely that the reason in all three cases is incomplete inter-mode thermal equilibration.

We start with a discussion of the physics at $v = 2/3$. Here, one *upstream* mode coexists with one *downstream* mode. They emanate from different reservoirs and can have different temperatures $T_u$ and $T_d$. In the absence of interactions, the heat current through each channel is

$$J(T) = \frac{\pi^2 k_B^2 T^2}{6h} \quad (17)$$

where $T_u$ and $T_d$ should be substituted in place of *T* for the *upstream* and *downstream* channels. Electron interactions result in energy exchange between the channels and make their temperatures coordinate-dependent. In the simplest model, the inter-channel heat current per unit length can be expressed in the spirit of Newton's law of cooling,

$$j_{ud} = \frac{\pi^2 k_B^2 (T_d^2 - T_u^2)}{6h\xi}, \qquad (18)$$

where $\xi$ is the temperature-equilibration length. The energy balance equation becomes

$$\frac{dJ(T_u[x])}{dx} = \frac{dJ(T_d[x])}{dx} = -j_{ud}[x] \ . \qquad (19)$$

Its solution reveals that the observed thermal conductance is diffusive, and thus far from the naïve zero value at full equilibrium (at the sample length $L \gg \xi$). It takes the form

$$\kappa \sim \frac{1}{1+L/\xi} \ . \qquad (20)$$

Detailed analysis [140, 144] shows that the equilibration length increases at lower temperatures, and hence, the thermal conductance grows as the temperature decreases – as observed.

The above discussion crucially depends on the equal numbers of the *upstream* and *downstream* channels at $\nu = 2/3$. In the states with an *unequal* numbers of *up-* and *down-stream* channels, finite size corrections to the equilibrium thermal conductance are much smaller than in Eq. (20) and exhibit an exponential dependence on $L/\xi$ [139, 140]. At first sight, this makes puzzling the observed deviations from the expected equilibrated thermal conductance at $\nu = 8/3$. The observed behavior can be explained by decoupling of some of the channels from the rest of the system [144]. Strong Coulomb interaction between integer and fractional edge channels makes it meaningless to assign different temperatures to integer and fractional channels. The right language to use is 'weakly vs strongly interacting modes'. In this case, the appropriate weakly interacting modes are the overall *downstream* charged mode of the conductance $\frac{8e^2}{3h}$, the spin mode, the upstream neutral mode, and one more neutral downstream mode [144]. Since the *downstream* charged mode is much faster than the neutral modes, the thermal excitations of that mode rapidly travel through the system before they can exchange their energy with the neutral modes. This example gives a hint of the importance of equilibration among modes in thermal measurements.

The observed increase of the thermal conductance at low temperatures at $\nu = 5/2$ cannot be easily explained in the same way. With the assumption that the state is a *PH-Pfaffian* liquid, it is crucial that the PH-Pfaffian equilibration length diverges faster at low temperatures than it does in the other possible states [140]. A lack of equilibration was also used to explain the observed thermal conductance at $\nu = 5/2$ assuming the *anti-Pfaffian* order, however, these explanations face difficulties [144-147].

## 8. Conclusion

In the first twenty years of the quantum Hall effect the focus of the experimental research was on bulk physics. In the late 1990s, the focus began shifting towards probing the edges. The first major success in the fractional regime was the detection of fractional charges via shot noise [34, 35], followed by the Nobel Prize to Stormer, Tsui, and Laughlin in 1998. Experiments concentrated in the following years on the properties of exotic states that harbor multiple and counter-propagating edge modes. Hand in hand, material quality improved substantially, mainly in GaAs-AlGaAs heterojunctions [143, 148, 149]. Follow up work has concentrated on deducing bulk topological orders from edge behavior.

One major research direction has been QHE interferometry, which may serve as a beacon to the anyonic statistics, be it abelian or non-abelian. Fabry-Perot (FPI) [66-69] and Mach-Zehnder (MZI) [64, 65] interferometers were employed. The FPI can be constructed in smaller dimensions, yet, it tends to suffer from strong Coulomb interaction, which hides the Aharonov-Bohm (AB) interference - making the interpretation of the data more difficult. The MZI, on the other hand, being larger (harboring an ohmic contact in its interior), shows always a non-interacting, two-path, AB interference. Interference in the integer regime is relatively easy; yet, new and unexpected effects have been discovered (electron pairing [84] and lobe-structure of the visibility [65]) – likely due to interactions in the integer regime. These unexpected effects prove the absence of full understanding even in the integer regime. On the other hand, experimental evidence of AB interference in the fractional regime is rather scarce, preventing thus far an assured demonstration of the anyonic nature of the quasiparticles [86, 126, 150]. Neutral modes (either topological or due to spontaneous edge-reconstruction) may be responsible for the dephasing in the fractional regime [63].

*Upstream* neutral edge modes, though theoretically predicted earlier [50], were observed only more recently [52-57]. Indeed, measuring energy transport is more difficult than measuring the transport of charge. The results gave a strong confirmation of the existing theory of abelian topological orders. Even more interesting results were obtained for the $\nu$=5/2 state, where a non-abelian topological order has long been suspected.

Numerical works have produced preponderance of evidence in favor of the non-abelian *Pfaffian* and *anti-Pfaffian* orders [116-118]. At the same time, experimental evidence seems to point towards a related, but distinct, *PH-Pfaffian* order [107]. A recent thermal conductance experiment [140], performed in an extremely high-mobility 2DEG buried in GaAs-AlGaAs heterostructures, showed results consistent with the *PH-Pfaffian* order. If one accepts the *PH-Pfaffian* order, then what is wrong with the numerical prediction?

Numerical works treat Landau level mixing (LLM) in an approximate manner, and it was suggested that LLM may be responsible for a different topological order in realistic samples [151]. More attention has focused on disorder-based explanations [107, 152, 153]. Indeed, until a recent preprint [154], no attempts were made to include disorder in simulations of the 5/2 physics. Under appropriate conditions, LLM and disorder may result in the macroscopic *PH-Pfaffian* order. It was proposed that a system with long correlations in its disorder can split into microscopic domains of Pfaffian and anti-Pfaffian liquids with a *PH-Pfaffian* emerging as a coherent average [152, 153]. A different approach was taken by interpreting the data in terms of the *anti-Pfaffian* state, assuming partial equilibration [144]. Yet, this interpretation must satisfy a few stringent constraints [146].

One thing is clear: whatever future developments await us, edge probes of topological orders will play a crucial role.

**Acknowledgements**

MH acknowledges W. Yang and V. Umansky for their extremely valuable help with the manuscript; the partial support of the Israeli Science Foundation (ISF), the Minerva foundation, and the European Research Council under the European Community's Seventh Framework Program (FP7/2007– 2013)/ERC Grant agreement 339070. DEF acknowledges partial support of the NSF under grant No. DMR-1607451.

**Appendix A. Edge Actions**

**A.1.** $\nu = 1$

We briefly address the origin of the action in Eq. (3):

$$\frac{S}{\hbar} = -\frac{1}{4\pi}\int dx\, dt[\partial_t\varphi\partial_x\varphi + v(\partial_x\varphi)^2] \qquad (A1)$$

The second term in the action, quadratic in $\partial_x\varphi$, is the energy cost of changing the charge density away from the minimal-energy configuration. To justify the first term and the coefficient $v$ in front of the second term, we make two observations: (*i*) The equation of motion for the action,

$$\partial_t\partial_x\varphi + v\partial_x^2\varphi = 0 \qquad (A2)$$

reveals chiral transport with the speed $v$, exactly as in the fermionic formulation; (*ii*) The action gives the correct conductance. One can check it by adding the contribution of the chemical potential $\int dx dt \rho V$, minimizing the action with respect to the charge density, and finally computing the current $j = ev\partial_x\varphi/2\pi$ from the equation of motion. The structure of the action determines the commutation relations for the field $\varphi$. Indeed, the action combines the Hamiltonian with a piece, containing the time derivative of $\varphi$. This piece tells us what variable is canonically conjugated to $\varphi$. The corresponding commutation relation is $[\varphi(x), \partial_y\varphi(y)] = -2\pi i\delta(x-y)$, or equivalently, $[\varphi(x), \varphi(y)] = \pi i\,\text{sign}(x-y)$. This allows identifying the electron operator $\psi$, which is a Fermi operator that changes the total charge $Q = \int dx\rho$ by an electron charge $e$. The above commutation relations shows that any operator of the form $\exp(i[2n+1]\varphi)$ satisfies the Fermi anti-commutation relation. Finally, the effect of electron annihilation on the charge of the edge implies the commutation relation $[\psi, Q] = e\psi$, being consistent with $\psi(x) = \exp(i\varphi(x))$.

**A.2.** *A general Abelian edge*

A systematic way to construct edge actions is known as the *K*-matrix formalism [3]. The topological properties of the bulk are encoded by an integer symmetric matrix *K* and an integer column *t*. Each bulk excitation corresponds to an integer column $l_k$. The filling factor equals $t^T K^{-1} t$, the quasiparticle charge is $e l_k^T K^{-1} t$, and the statistical phase accumulated by an anyon of type $l_k$ making a full circle around an anyon of type $l_m$ is $2\pi l_k^T K^{-1} l_m$. Unimodular (determinant=1) integer matrices *W* define equivalent descriptions of the same phase with $K \to WKW^T$, $t \to Wt$. According to the 'bulk-edge correspondence' principle, topological properties of the edge are determined by those of the bulk. In particular, the simplest edge theory is given by the action

$$\frac{S}{\hbar} = -\frac{1}{4\pi} \int dx\, dt \sum_{km} [K_{km}\, \partial_t \varphi_k \partial_x \varphi_m + V_{km} \partial_x \varphi_k \partial_x \varphi_m] \,, \quad (A3)$$

where the first term contains the bulk $K$-matrix and the second term $V$ describes interactions. To identify edge channels one needs to diagonalize simultaneously the symmetric matrices $K$ and $V$. The signs of the eigenvalues of $K$ determine the propagation directions of the modes. One can deduce the $K$ matrices at $\nu = 1$, 1/3, and 2/3 from the discussion above and in Section 2.

### A.3. Disorder on the edge at $\nu = 2/3$

The simplest edge model of the hole-conjugate $\nu = 2/3$ state has the action

$$\frac{S}{\hbar} = -\frac{1}{4\pi} \int dx\, dt [\partial_t \varphi_1 \partial_x \varphi_1 + v_1 (\partial_x \varphi_1)^2]$$
$$+ \frac{3}{4\pi} \int dx\, dt [\partial_t \varphi_{1/3} \partial_x \varphi_{1/3} - v_{1/3} (\partial_x \varphi_{1/3})^2] \quad (A4)$$

where the fields $\varphi_1$ and $\varphi_{1/3}$ define the charge densities within the two channels. As discussed in Section 4, such an action results in an incorrect electrical conductance. The goal of this section is to incorporate the missing inter-channel interaction and tunneling into the model. This allows fixing the conductance issue.

The new model represents the total charge density at the edge as $e\partial_x \varphi_c / 2\pi = e\partial_x (\varphi_1 + \varphi_{1/3})/2\pi$, accompanied by a neutral mode $\varphi_n = (\varphi_1 + 3\varphi_{1/3})/\sqrt{2}$. The form of $\varphi_n$ is motivated by: (*i*) This choice simplifies the expression for the tunneling operator that transfers an electron between the two edge channels: $T = \exp(\pm\sqrt{2}i\varphi_n)$; and (*ii*) The $K$-matrix contribution to the action in Eq. A3 remains diagonal in the variables $\varphi_n$ and $\varphi_c$. The action than becomes,

$$\frac{S}{\hbar} = -\int dxdt \left\{ \frac{1}{4\pi}[-\partial_t \varphi_n \partial_x \varphi_n + v_n(\partial_x \varphi_n)^2] + \frac{3}{8\pi}[\partial_t \varphi_c \partial_x \varphi_c + v_c(\partial_x \varphi_c)^2] + \frac{w_{nc}}{4\pi} \partial_x \varphi_n \partial_x \varphi_c + [W(x)\exp(\sqrt{2}i\varphi_n) + H.c.] \right\}, \quad (A5)$$

where $w_{nc}$ describes the interaction between the neutral and charged modes, and $W(x)$ is the position-dependent tunneling amplitude.

The mode velocities $v_c$ and $v_n$ are determined by the Coulomb interaction, similarly to $w_{nc}$, and it is expected that $v_c > v_n, w_{nc}$. On an etched edge, where the Coloumb interaction is not screened by defining gates, $v_c$ is dominated by the long-range part of the Coloumb force, and the velocity of the low-energy excitations of the charge mode diverges logarithmically with the wave-length.

$W(x)$ depends on the microscopic details. In a clean sample, $W(x) \sim \exp(ix\Delta k)$ carries information about the momentum mismatch $\Delta k$ of the edge modes $\varphi_1$ and $\varphi_{1/3}$. Indeed, Fig. 2 shows that these modes are spatially separated, while Eq. (1) shows that spatial separation translates into momentum difference in a strong perpendicular magnetic field. This leads us to a new difficulty: in the low-energy limit, no tunneling processes satisfy both energy and momentum conservation. Thus, charge tunneling is ineffective and we still get a wrong conductance. The crucial observation is the role of disorder [50], since in any realistic sample, $W(x)$ is a random function of the coordinate.

The role of the random tunneling term can be understood from its behavior under *renormalization group* (RG) transformations [49, 50]. In the lowest order, the RG procedure consists of two steps: (*i*) Using the quadratic part of the action (Eq. (A5)) to integrate out the large wave-vector part of $\varphi_n(x)$ from the tunneling operator $\exp(\sqrt{2}i\varphi_n)$; and (*ii*) Rescaling *x* and *t*. For a coordinate-independent $W(x)$, the rescaling of *x* and *t* by a factor of *s* would multiply $W$ by $s^2$. Since $W$ is spatially random, the rescaling of *x* generates a factor of $\sqrt{s}$, so that the total rescaling factor is $s^{3/2}$. Assuming $v_c \gg w_{nc}$, one finds that the renormalized tunneling amplitude grows as $W \sim l^{1/2}$, where the length *l* is the running RG ultraviolet cut-off scale. Thus, even arbitrarily small $W$ results in strong inter-channel tunneling at low energies on a sufficiently long edge.

Since RG flows into the strong coupling regime, it is not obvious, what exactly happens at low energies. Kane, Fisher, and Polchiski [50] found an answer by fermionizing the action (Eq. (A5)), leading to decoupled charge and neutral modes:

$$\frac{S}{\hbar} = -\int dx dt \left\{ \frac{1}{4\pi}[-\partial_t \varphi_n \partial_x \varphi_n + v_n(\partial_x \varphi_n)^2] + \frac{3}{8\pi}[\partial_t \varphi_c \partial_x \varphi_c + v_c(\partial_x \varphi_c)^2] \right\} \quad (A6)$$

This action describes the RG fixed point. In general, irrelevant perturbations are also present.

**Appendix B.  Bulk-edge correspondence at $\nu = 5/2$.**

The goal of this section is to explain why an odd number of edge Majorana modes corresponds to non-Abelian statistics in the bulk, and an even number of edge Majorana modes corresponds to Abelian statistics (Section 6.2). Our approach is heuristic. A more rigorous discussion is beyond the scope of this chapter. Our discussion builds on *bulk-edge* correspondence.

We interpret the bulk anyons as small holes in a 2D electron gas, with each hole having a perimeter and hence carrying edge modes with an action of the type of Eq. (14). Such picture does not reflect the microscopic details of realistic samples, since propagation of these holes would involve dramatic destruction and reconstruction of chemical bonds in the material. Still, it is sensible to expect that this simplistic model captures the universal topological properties. Besides, the 2D gas can be locally depleted by a scanning tip. The resulting "hole" in the gas can bind anyons.

Let us start with the Chern number $\nu_C = 2$. The two neutral edge modes around the hole can be combined into a single Dirac fermion. For a large hole, the edge spectrum is almost continuous. Yet, in the limit of a small hole, most edge states have a high energy and drop out from low-energy physics. In the simplest model, exactly one fermionic state survives in the low-energy limit. The state can be either filled or empty. This corresponds to an almost trivial Hamiltonian $H = \varepsilon \Psi^+ \Psi$, where $\Psi$ is a complex fermion operator, represented as a combination of two real fermions *a* and *b*: $\Psi = a + ib$. Since $a^2$ and $b^2$ are *c*-numbers, the Hamiltonian reduces to,

$$H = 2i\varepsilon ab + const. \tag{B1}$$

Effectively, two neutral Majorana fermions live in the hole. With two holes in close proximity in a $\nu_C = 1$ system and each hole carrying a single Majorana mode, there are two of them altogether. Consequently, it is sensible to expect that the low-energy physics is similar to that of a single hole at $\nu_C = 2$, and that the Hamiltonian is still the one in Eq. (B1). When one moves the two holes far from each other, the general structure of the Hamiltonian remains; however, the distant neutral excitations cannot interact anymore (thus, $\varepsilon \to 0$). The system hence possesses two degenerate states of different total fermionic parity $(a - ib)(a + ib)$. In other words, the two holes can form states of two different statistics. The holes are anyons in our model, and so we have discovered two fusion possibilities.

To establish non-Abelian statistics, each hole should represent the same anyon type irrespectively of the fusion channel. If this were not the case, some local observable would be different in the two fusion channels. Local observables are Hermitian Bose-operators built from *a* only or from *b* only. Any such operator is an even power of a Majorana fermion, and hence, a trivial constant. In other words, no local measurements can distinguish the states of the holes in the two fusion channels, and both fusion outcomes emerge from a combination of the same anyons.

**Appendix C.   Coulomb dominated regime in the FQHE regime**

For $f_T=0$ in the integer regime, increasing $B$ leads to area shrinking, keeping the threading flux constant, and to imbalance between electrons and ionized donors. Eventually, the imbalance relaxes (when interactions reach the charging energy), by adding a quasiparticle within the interfering Landau level. The area returns to its initial state with the addition of a flux quantum - being invisible in the interference pattern. For $f_T>0$, with increasing $B$ the interfering channel loses electrons to the $f_T$ lower Landau levels (whose density increases with field), and the threading flux reduces (opposite to the behavior in the pure AB regime).

We discuss now the periodicity in the gate voltage $\Delta V_{MG}$, as it provides an insight to the charge of the interfering quasiparticle. We assume that the capacitance $C$ between the modulation gate and the interfering channel depends only on $f_T$, namely, $C=C(f_T)$. The interfering channel flows at the interface between two areas with different filling factors, $\nu_{out}$ and $\nu_{in}$, each with quasiparticle charge $q_i=e\nu_i$. As long as the sole function of biasing the plunger gate by $\delta V_{MG}$ is to move the interface between the two filling factors by area $\delta A$, and thus expel a $\delta q$ charge from within the interferometer, we may write:

$$\delta q = C\delta V_{MG} = \frac{B\cdot\delta A}{\phi_0}\cdot e(\nu_{in}-\nu_{out}) \tag{C1}$$

Assuming that the change in the area does not change the number of quasiparticles enclosed by the interfering loop (large energy is required to induce quasiparticles or quasiholes), the change of the area leads to a change in the AB phase,

$$\delta\varphi = 2\pi\frac{e^*}{e}\frac{B\cdot\delta A}{\phi_0} \tag{C2}$$

Combining Eq. (C1) and (C2) we get a relation between the interfering quasiparticle charge $e^*$ and the periodicity $\Delta V_{MG}$:

$$\frac{e^*}{e} = \frac{\nu_{in}-\nu_{out}}{\Delta V_{MG}/\Delta V_e}, \tag{C3}$$

where $\Delta V_e$ is the gate voltage needed to expel one electron.

Indeed, in the cases (1/3, 0) and (2/5, 0), the periodicities in the gate voltage are the same as for $f_T$=0 in the integer cases; namely, the expelled charge per period must be in both cases $e$. Since the interfering edge channel in both cases belongs to the 1/3 fractional state, $v_{in}$-$v_{out}$=1/3, the interfering quasiparticle charge must be $e^*=e/3$ (Eq. (C3)) [69].

In the case (2/5, 1/3), where the interfering channel belongs to the 2/5 fractional state, the observed [69] periodicity is nearly 1/3 of the period in the integer cases, and thus the expelled charge must be $e/3$. As $v_{in}$-$v_{out}$=1/15 the interfering charge is $e^*=e/5$. This is a striking example of an expelled charge $e/3$ per period of gate voltage, while the interfering quasiparticles carried charge $e/5$ [69].

**References**

1. B.I. Halperin, Quantized Hall conductance, current-carrying edge states, and the existence of extended states in a two-dimensional disordered potential, *Phys. Rev. B*, **25**(4), 2185-2190, (1982).
2. X.G. Wen, Chiral Luttinger liquid and the edge excitations in the fractional quantum Hall states, *Phys. Rev. B*, **41**(18), 12838-12844, (1990).
3. X.-G. Wen, Quantum field theory of many-body systems: From the origin of sound to an origin of light and electrons, (Oxford University Press on Demand 2004).
4. C.L. Kane, M.P.A. Fisher, Edge-State Transport, in: S. Das Sarma, A. Pinczuk (Eds.) *Perspectives in Quantum Hall Effects: Novel Quantum Liquids in Low‐Dimensional Semiconductor Structures*, (John Wiley, New York, 1996).
5. M. Buttiker, The quantum Hall-effect in open conductors, in: M. Reed (Ed.) *Semiconductors and Semimetals, Vol 35: Nanostructured Systems*, (Elsevier Academic Press Inc, San Diego, 1992).
6. D.B. Chklovskii, B.I. Shklovskii, L.I. Glazman, Electrostatics of edge channels, *Phys. Rev. B*, **46**(7), 4026-4034, (1992).
7. B.Y. Gelfand, B.I. Halperin, Edge electrostatics of a mesa-etched sample and edge-state-to-bulk scattering rate in the fractional quantum Hall regime, *Phys. Rev. B*, **49**(3), 1862-1866, (1994).
8. N. Pascher, C. Rossler, T. Ihn, K. Ensslin, C. Reichl, W. Wegscheider, Imaging the conductance of integer and fractional quantum Hall edge states, *Phys. Rev. X*, **4**(1), 011014, (2014).
9. C. Chamon, X.G. Wen, Sharp and smooth boundaries of quantum Hall liquids, *Phys. Rev. B*, **49**(12), 8227-8241, (1994).
10. H. Inoue, A. Grivnin, Y. Ronen, M. Heiblum, V. Umansky, D. Mahalu, Proliferation of neutral modes in fractional quantum Hall states, *Nat. Comm.*, **5**(4067, (2014).
11. J.K. Jain, Composite-fermion approach for the fractional quantum Hall effect, *Phys. Rev. Lett.*, **63**(2), 199-202, (1989).
12. J.K. Jain, S.A. Kivelson, D.J. Thouless, Proposed measurement of an effective flux quantum in the fractional quantum Hall effect, *Phys. Rev. Lett.*, **71**(18), 3003-3006, (1993).
13. A.H. MacDonald, Edge states in the fractional-quantum-Hall-effect regime, *Phys. Rev. Lett.*, **64**(2), 220-223, (1990).
14. Y. Cohen, Y. Ronen, W. Yang, D. Banitt, J. Park, M. Heiblum, A.D. Mirlin, Y. Gefen, V. Umansky, Synthesizing a $v = 2/3$ fractional quantum Hall effect edge state from counter-propagating $v = 1$ and $v = 1/3$ states, *Nat. Comm.*, **10**(1920, (2019).
15. Y. Meir, Composite edge states in the $v = 2/3$ fractional quantum Hall regime, *Phys. Rev. Lett.*, **72**(16), 2624-2627, (1994).
16. R. Sabo, I. Gurman, A. Rosenblatt, F. Lafont, D. Banitt, J. Park, M. Heiblum, Y. Gefen, V. Umansky, D. Mahalu, Edge reconstruction in fractional quantum Hall states, *Nat. Phys.*, **13**(5), 491-496, (2017).
17. R.B. Laughlin, Anomalous quantum Hall effect - an incompressible quantum fluid with fractionally charged excitations, *Phys. Rev. Lett.*, **50**(18), 1395-1398, (1983).
18. M. Levin, A. Stern, Fractional topological insulators, *Phys. Rev. Lett.*, **103**(19), 196803, (2009).
19. Y.M. Blanter, M. Buttiker, Shot noise in mesoscopic conductors, *Phys. Rep.*, **336**(1-2), 1-166, (2000).
20. W. Schottky, Über spontane stromschwankungen in verschiedenen elektrizitätsleitern, *Ann. Phys.*, **362**(23), 541-567, (1918).

21. W. Schottky, On spontaneous current fluctuations in various electrical conductors, *J. Micro. Nanolith. MEM.* , **17**(4), 1-11, (2018).
22. C. Chamon, D.E. Freed, X.G. Wen, Tunneling and quantum noise in one-dimensional Luttinger liquids, *Phys. Rev. B*, **51**(4), 2363-2379, (1995).
23. C.L. Kane, M.P. Fisher, Nonequilibrium noise and fractional charge in the quantum Hall effect, *Phys. Rev. Lett.*, **72**(5), 724-727, (1994).
24. T. Martin, R. Landauer, Wave-packet approach to noise in multichannel mesoscopic systems, *Phys. Rev. B*, **45**(4), 1742-1755, (1992).
25. G.B. Lesovik, Excess quantum noise in 2d ballistic point contacts, *J. Exp. Theor. Phys*, **49**(9), 592-594, (1989).
26. V. A Khlus, Current and voltage fluctuations in microjunctions between normal metals and superconductors, *J. Exp. Theor. Phys*, **66**(6), 2179, (1987).
27. M. Reznikov, M. Heiblum, H. Shtrikman, D. Mahalu, Temporal correlation of electrons: Suppression of shot noise in a ballistic quantum point contact, *Phys. Rev. Lett.*, **75**(18), 3340-3343, (1995).
28. A. Kumar, L. Saminadayar, D.C. Glattli, Y. Jin, B. Etienne, Experimental test of the quantum shot noise reduction theory, *Phys. Rev. Lett.*, **76**(15), 2778-2781, (1996).
29. M. Reznikov, E. De Picciotto, M. Heiblum, D.C. Glattli, A. Kumar, L. Saminadayar, Quantum shot noise, *Superlattice Microst.*, **23**(3-4), 901-915, (1998).
30. S. Ilani, J. Martin, E. Teitelbaum, J.H. Smet, D. Mahalu, V. Umansky, A. Yacoby, The microscopic nature of localization in the quantum Hall effect, *Nature*, **427**(6972), 328-332, (2004).
31. J.A. Simmons, S.W. Hwang, D.C. Tsui, H.P. Wei, L.W. Engel, M. Shayegan, Resistance fluctuations in the integral-quantum-Hall-effect and fractional-quantum-Hall-effect regimes, *Phys. Rev. B*, **44**(23), 12933-12944, (1991).
32. V.J. Goldman, B. Su, Resonant tunneling in the quantum Hall regime: Measurement of fractional charge, *Science*, **267**(5200), 1010-1012, (1995).
33. J.D.F. Franklin, I. Zailer, C.J.B. Ford, P.J. Simpson, J.E.F. Frost, D.A. Ritchie, M.Y. Simmons, M. Pepper, The Aharonov-Bohm effect in the fractional quantum Hall regime, *Surf. Sci.*, **361**(1-3), 17-21, (1996).
34. R. dePicciotto, M. Reznikov, M. Heiblum, V. Umansky, G. Bunin, D. Mahalu, Direct observation of a fractional charge, *Nature*, **389**(6647), 162-164, (1997).
35. L. Saminadayar, D.C. Glattli, Y. Jin, B. Etienne, Observation of the e/3 fractionally charged Laughlin quasiparticle, *Phys. Rev. Lett.*, **79**(13), 2526-2529, (1997).
36. Y.C. Chung, M. Heiblum, V. Umansky, Scattering of bunched fractionally charged quasiparticles, *Phys. Rev. Lett.*, **91**(21), 216804, (2003).
37. Y.C. Chung, M. Heiblum, Y. Oreg, V. Umansky, D. Mahalu, Anomalous chiral Luttinger liquid behavior of diluted fractionally charged quasiparticles, *Phys. Rev. B*, **67**(20), 201104, (2003).
38. E. Comforti, Y.C. Chung, M. Heiblum, A.V. Umansky, Multiple scattering of fractionally charged quasiparticles, *Phys. Rev. Lett.*, **89**(6), 066803, (2002).
39. E. Comforti, Y.C. Chung, M. Heiblum, V. Umansky, D. Mahalu, Bunching of fractionally charged quasiparticles tunnelling through high-potential barriers, *Nature*, **416**(6880), 515-518, (2002).
40. P. Fendley, A.W. Ludwig, H. Saleur, Exact nonequilibrium dc shot noise in Luttinger liquids and fractional quantum Hall devices, *Phys. Rev. Lett.*, **75**(11), 2196-2199, (1995).
41. P. Fendley, A.W. Ludwig, H. Saleur, Exact conductance through point contacts in the $v = 1/3$ fractional quantum Hall effect, *Phys. Rev. Lett.*, **74**(15), 3005-3008, (1995).

42. P. Fendley, A.W.W. Ludwig, H. Saleur, Exact nonequilibrium transport through point contacts in quantum wires and fractional quantum Hall devices, *Phys. Rev. B*, **52**(12), 8934-8950, (1995).
43. P. Fendley, H. Saleur, Nonequilibrium de noise in a Luttinger liquid with an impurity, *Phys. Rev. B*, **54**(15), 10845-10854, (1996).
44. T.G. Griffiths, E. Comforti, M. Heiblum, A. Stern, V.V. Umansky, Evolution of quasiparticle charge in the fractional quantum hall regime, *Phys. Rev. Lett.*, **85**(18), 3918-3921, (2000).
45. C.J. Wang, D.E. Feldman, Fluctuation-dissipation theorem for chiral systems in nonequilibrium steady states, *Phys. Rev. B*, **84**(23), 235315, (2011).
46. C. Wang, D.E. Feldman, Chirality, causality, and fluctuation-dissipation theorems in nonequilibrium steady states, *Phys. Rev. Lett.*, **110**(3), 030602, (2013).
47. C.J. Wang, D.E. Feldman, Fluctuation theorems without time-reversal symmetry, *Int. J. Mod. Phys. B*, **28**(7), 1430003, (2014).
48. D.E. Feldman, M. Heiblum, Why a noninteracting model works for shot noise in fractional charge experiments, *Phys. Rev. B*, **95**(11), 115308, (2017).
49. C.L. Kane, M.P. Fisher, Impurity scattering and transport of fractional quantum Hall edge states, *Phys. Rev. B*, **51**(19), 13449-13466, (1995).
50. C.L. Kane, M.P. Fisher, J. Polchinski, Randomness at the edge: Theory of quantum Hall transport at filling $\nu = 2/3$, *Phys. Rev. Lett.*, **72**(26), 4129-4132, (1994).
51. D.E. Feldman, F.F. Li, Charge-statistics separation and probing non-Abelian states, *Phys. Rev. B*, **78**(16), 161304, (2008).
52. A. Bid, N. Ofek, H. Inoue, M. Heiblum, C.L. Kane, V. Umansky, D. Mahalu, Observation of neutral modes in the fractional quantum Hall regime, *Nature*, **466**(7306), 585-590, (2010).
53. Y. Gross, M. Dolev, M. Heiblum, V. Umansky, D. Mahalu, Upstream neutral modes in the fractional quantum Hall effect regime: Heat waves or coherent dipoles, *Phys. Rev. Lett.*, **108**(22), 226801, (2012).
54. V. Venkatachalam, S. Hart, L. Pfeiffer, K. West, A. Yacoby, Local thermometry of neutral modes on the quantum Hall edge, *Nat. Phys.*, **8**(9), 676-681, (2012).
55. M. Dolev, Y. Gross, R. Sabo, I. Gurman, M. Heiblum, V. Umansky, D. Mahalu, Characterizing neutral modes of fractional states in the second Landau level, *Phys. Rev. Lett.*, **107**(3), 036805, (2011).
56. I. Gurman, R. Sabo, M. Heiblum, V. Umansky, D. Mahalu, Extracting net current from an upstream neutral mode in the fractional quantum Hall regime, *Nat. Comm.*, **3**(1289, (2012).
57. A. Rosenblatt, F. Lafont, I. Levkivskyi, R. Sabo, I. Gurman, D. Banitt, M. Heiblum, V. Umansky, Transmission of heat modes across a potential barrier, *Nat. Comm.*, **8**(1), 2251, (2017).
58. M. Levin, B.I. Halperin, B. Rosenow, Particle-hole symmetry and the Pfaffian state, *Phys. Rev. Lett.*, **99**(23), 236806, (2007).
59. B.J. Overbosch, C. Chamon, Long tunneling contact as a probe of fractional quantum Hall neutral edge modes, *Phys. Rev. B*, **80**(3), 035319, (2009).
60. E. Grosfeld, S. Das, Probing the neutral edge modes in transport across a point contact via thermal effects in the Read-Rezayi non-Abelian quantum Hall states, *Phys. Rev. Lett.*, **102**(10), 106403, (2009).
61. U. Klass, W. Dietsche, K. Vonklitzing, K. Ploog, Imaging of the dissipation in quantum-Hall effect experiments, *Z. Phys. B Con. Mat.*, **82**(3), 351-354, (1991).
62. C. Spånslätt, J. Park, Y. Gefen, A. Mirlin, *Topological classification of shot noise on fractional quantum Hall edges*, arXiv:1906.05623, (2019).

63. R. Bhattacharyya, M. Banerjee, M. Heiblum, D. Mahalu, V. Umansky, Melting of interference in the fractional quantum Hall effect: Appearance of neutral modes, *Phys. Rev. Lett.*, **122**(24), 246801, (2019).
64. Y. Ji, Y. Chung, D. Sprinzak, M. Heiblum, D. Mahalu, H. Shtrikman, An electronic Mach-Zehnder interferometer, *Nature*, **422**(6930), 415-418, (2003).
65. I. Neder, M. Heiblum, Y. Levinson, D. Mahalu, V. Umansky, Unexpected behavior in a two-path electron interferometer, *Phys. Rev. Lett.*, **96**(1), 016804, (2006).
66. C.D.C. Chamon, D.E. Freed, S.A. Kivelson, S.L. Sondhi, X.G. Wen, Two point-contact interferometer for quantum Hall systems, *Phys. Rev. B*, **55**(4), 2331-2343, (1997).
67. J.A. Folk, C.M. Marcus, J.S. Harris, Jr., Decoherence in nearly isolated quantum dots, *Phys. Rev. Lett.*, **87**(20), 206802, (2001).
68. A. Kou, C.M. Marcus, L.N. Pfeiffer, K.W. West, Coulomb oscillations in antidots in the integer and fractional quantum Hall regimes, *Phys. Rev. Lett.*, **108**(25), 256803, (2012).
69. N. Ofek, A. Bid, M. Heiblum, A. Stern, V. Umansky, D. Mahalu, Role of interactions in an electronic Fabry–Perot interferometer operating in the quantum Hall effect regime, *Proc. Natl. Acad. Sci.*, **107**(12), 5276-5281, (2010).
70. P. Roulleau, F. Portier, D.C. Glattli, P. Roche, A. Cavanna, G. Faini, U. Gennser, D. Mailly, Finite bias visibility of the electronic Mach-Zehnder interferometer, *Phys. Rev. B*, **76**(16), 161309, (2007).
71. L.V. Litvin, A. Helzel, H.P. Tranitz, W. Wegscheider, C. Strunk, Edge-channel interference controlled by Landau level filling, *Phys. Rev. B*, **78**(7), 075303, (2008).
72. E. Bieri, M. Weiss, O. Goktas, M. Hauser, C. Schonenberger, S. Oberholzer, Finite-bias visibility dependence in an electronic Mach-Zehnder interferometer, *Phys. Rev. B*, **79**(24), 245324, (2009).
73. I. Neder, M. Heiblum, D. Mahalu, V. Umansky, Entanglement, dephasing, and phase recovery via cross-correlation measurements of electrons, *Phys. Rev. Lett.*, **98**(3), 036803, (2007).
74. I. Neder, F. Marquardt, M. Heiblum, D. Mahalu, V. Umansky, Controlled dephasing of electrons by non-gaussian shot noise, *Nat Phys*, **3**(8), 534-537, (2007).
75. P. Roulleau, F. Portier, P. Roche, A. Cavanna, G. Faini, U. Gennser, D. Mailly, Tuning decoherence with a voltage probe, *Phys. Rev. Lett.*, **102**(23), 236802, (2009).
76. P. Roulleau, F. Portier, P. Roche, A. Cavanna, G. Faini, U. Gennser, D. Mailly, Noise dephasing in edge states of the integer quantum Hall regime, *Phys. Rev. Lett.*, **101**(18), 186803, (2008).
77. H. Inoue, A. Grivnin, N. Ofek, I. Neder, M. Heiblum, V. Umansky, D. Mahalu, Charge fractionalization in the integer quantum Hall effect, *Phys. Rev. Lett.*, **112**(16), 166801, (2014).
78. A. Helzel, L.V. Litvin, I.P. Levkivskyi, E.V. Sukhorukov, W. Wegscheider, C. Strunk, Counting statistics and dephasing transition in an electronic Mach-Zehnder interferometer, *Phys. Rev. B*, **91**(24), 245419, (2015).
79. I.P. Levkivskyi, E.V. Sukhorukov, Dephasing in the electronic Mach-Zehnder interferometer at filling factor $v = 2$, *Phys. Rev. B*, **78**(4), 045322, (2008).
80. M.J. Rufino, D.L. Kovrizhin, J.T. Chalker, Solution of a model for the two-channel electronic Mach-Zehnder interferometer, *Phys. Rev. B*, **87**(4), 045120, (2013).
81. D.L. Kovrizhin, J.T. Chalker, Multiparticle interference in electronic Mach-Zehnder interferometers, *Phys. Rev. B*, **81**(15), 155318, (2010).
82. B. Rosenow, B.I. Halperin, Influence of interactions on flux and back-gate period of quantum Hall interferometers, *Phys. Rev. Lett.*, **98**(10), 106801, (2007).

83. B.I. Halperin, A. Stern, I. Neder, B. Rosenow, Theory of the Fabry-Perot quantum Hall interferometer, *Phys. Rev. B*, **83**(15), 155440, (2011).
84. H.K. Choi, I. Sivan, A. Rosenblatt, M. Heiblum, V. Umansky, D. Mahalu, Robust electron pairing in the integer quantum hall effect regime, *Nat. Comm.*, **6**(7435, (2015).
85. Y.M. Zhang, D.T. McClure, E.M. Levenson-Falk, C.M. Marcus, L.N. Pfeiffer, K.W. West, Distinct signatures for Coulomb blockade and Aharonov-Bohm interference in electronic Fabry-Perot interferometers, *Phys. Rev. B*, **79**(24), 241304, (2009).
86. J. Nakamura, S. Fallahi, H. Sahasrabudhe, R. Rahman, S. Liang, G.C. Gardner, M.J. Manfra, Aharonov–Bohm interference of fractional quantum Hall edge modes, *Nat. Phys.*, **15**(6), 563-569, (2019).
87. I. Sivan, R. Bhattacharyya, H.K. Choi, M. Heiblum, D.E. Feldman, D. Mahalu, V. Umansky, Interaction-induced interference in the integer quantum Hall effect, *Phys. Rev. B*, **97**(12), 125405, (2018).
88. I. Sivan, H.K. Choi, J. Park, A. Rosenblatt, Y. Gefen, D. Mahalu, V. Umansky, Observation of interaction-induced modulations of a quantum Hall liquid's area, *Nat. Comm.*, **7**(12184, (2016).
89. E. Sela, Y. Oreg, F. von Oppen, J. Koch, Fractional shot noise in the kondo regime, *Phys. Rev. Lett.*, **97**(8), 086601, (2006).
90. O. Zarchin, M. Zaffalon, M. Heiblum, D. Mahalu, V. Umansky, Two-electron bunching in transport through a quantum dot induced by Kondo correlations, *Phys. Rev. B*, **77**(24), 241303, (2008).
91. N. Byers, C.N. Yang, Theoretical considerations concerning quantized magnetic flux in superconducting cylinders, *Phys. Rev. Lett.*, **7**(2), 46, (1961).
92. K.T. Law, D.E. Feldman, Y. Gefen, Electronic Mach-Zehnder interferometer as a tool to probe fractional statistics, *Phys. Rev. B*, **74**(4), 045319, (2006).
93. G. Moore, N. Read, Nonabelions in the fractional quantum Hall effect, *Nucl. Phys. B*, **360**(2-3), 362-396, (1991).
94. C. Nayak, S.H. Simon, A. Stern, M. Freedman, S. Das Sarma, Non-Abelian anyons and topological quantum computation, *Rev. Mod. Phys.*, **80**(3), 1083-1159, (2008).
95. A. Kitaev, Anyons in an exactly solved model and beyond, *Ann. Phys-New York*, **321**(1), 2-111, (2006).
96. R.L. Willett, K.W. West, L.N. Pfeiffer, Experimental demonstration of Fermi surface effects at filling factor 5/2, *Phys. Rev. Lett.*, **88**(6), 066801, (2002).
97. M.S. Hossain, M.K. Ma, M.A. Mueed, L.N. Pfeiffer, K.W. West, K.W. Baldwin, M. Shayegan, Direct observation of composite fermions and their fully-spin-polarized Fermi sea near $v$=5/2, *Phys. Rev. Lett.*, **120**(25), 256601, (2018).
98. N. Read, D. Green, Paired states of fermions in two dimensions with breaking of parity and time-reversal symmetries and the fractional quantum Hall effect, *Phys. Rev. B*, **61**(15), 10267-10297, (2000).
99. M. Dolev, M. Heiblum, V. Umansky, A. Stern, D. Mahalu, Observation of a quarter of an electron charge at the $v$ = 5/2 quantum Hall state, **452**(7189), 829-834, (2008).
100. J.B. Miller, I.P. Radu, D.M. Zumbuhl, E.M. Levenson-Falk, M.A. Kastner, C.M. Marcus, L.N. Pfeiffer, K.W. West, Fractional quantum Hall effect in a quantum point contact at filling fraction 5/2, *Nat. Phys.*, **3**(8), 561-565, (2007).
101. M. Dolev, Y. Gross, Y.C. Chung, M. Heiblum, V. Umansky, D. Mahalu, Dependence of the tunneling quasiparticle charge determined via shot noise measurements on the tunneling barrier and energetics, *Phys. Rev. B*, **81**(16), 161303, (2010).

102. R.L. Willett, L.N. Pfeiffer, K.W. West, Measurement of filling factor 5/2 quasiparticle interference with observation of charge e/4 and e/2 period oscillations, *Proc. Natl. Acad. Sci.*, **106**(22), 8853-8858, (2009).
103. K.K.W. Ma, D.E. Feldman, The sixteenfold way and the quantum Hall effect at half-integer filling factors, *Phys. Rev. B*, **100**(3), 035302, (2019).
104. S.S. Lee, S. Ryu, C. Nayak, M.P. Fisher, Particle-hole symmetry and the $v = 5/2$ quantum Hall state, *Phys. Rev. Lett.*, **99**(23), 236807, (2007).
105. G. Yang, D.E. Feldman, Experimental constraints and a possible quantum Hall state at $v = 5/2$, *Phys. Rev. B*, **90**(16), 161306, (2014).
106. D.T. Son, Is the composite fermion a Dirac particle?, *Phys. Rev. X*, **5**(3), 031027, (2015).
107. P.T. Zucker, D.E. Feldman, Stabilization of the particle-hole pfaffian order by Landau-Level mixing and impurities that break particle-hole symmetry, *Phys. Rev. Lett.*, **117**(9), 096802, (2016).
108. L. Fidkowski, X. Chen, A. Vishwanath, Non-abelian topological order on the surface of a 3D topological superconductor from an exactly solved model, *Phys. Rev. X*, **3**(4), 041016, (2013).
109. P. Bonderson, C. Nayak, X.-L. Qi, A time-reversal invariant topological phase at the surface of a 3D topological insulator, *J. Stat. Mech-Theory E.*, **2013**(09), P09016, (2013).
110. B.J. Overbosch, X.G. Wen, *Phase transitions on the edge of the $v = 5/2$ Pfaffian and anti-Pfaffian quantum Hall state*, arXiv:0804.2087, (2008).
111. B.I. Halperin, Theory of the quantized Hall conductance, *Helv. Phys. Acta*, **56**(1-3), 75-102, (1983).
112. X.G. Wen, Non-Abelian statistics in the fractional quantum Hall states, *Phys. Rev. Lett.*, **66**(6), 802-805, (1991).
113. J.K. Jain, Incompressible quantum Hall states, *Phys. Rev. B*, **40**(11), 8079-8082, (1989).
114. F.D. Haldane, E.H. Rezayi, Spin-singlet wave function for the half-integral quantum Hall effect, *Phys. Rev. Lett.*, **60**(10), 956-959, (1988).
115. A.H. MacDonald, D. Yoshioka, S.M. Girvin, Comparison of models for the even-denominator fractional quantum Hall effect, *Phys. Rev. B*, **39**(11), 8044-8047, (1989).
116. R.H. Morf, Transition from quantum Hall to compressible states in the second Landau Level: new light on the v=5/2 enigma, *Phys. Rev. Lett.*, **80**(7), 1505-1508, (1998).
117. K. Pakrouski, M.R. Peterson, T. Jolicoeur, V.W. Scarola, C. Nayak, M. Troyer, Phase diagram of the $v = 5/2$ fractional quantum Hall effect: Effects of Landau-Level mixing and nonzero width, *Phys. Rev. X*, **5**(2), 021004, (2015).
118. E.H. Rezayi, Landau Level mixing and the ground state of the v=5/2 quantum Hall effect, *Phys. Rev. Lett.*, **119**(2), 026801, (2017).
119. S. Baer, C. Rossler, T. Ihn, K. Ensslin, C. Reichl, W. Wegscheider, Experimental probe of topological orders and edge excitations in the second Landau level, *Phys. Rev. B*, **90**(7), 075403, (2014).
120. I.P. Radu, J.B. Miller, C.M. Marcus, M.A. Kastner, L.N. Pfeiffer, K.W. West, Quasi-particle properties from tunneling in the $v = 5/2$ fractional quantum Hall state, *Science*, **320**(5878), 899-902, (2008).
121. X. Lin, C. Dillard, M.A. Kastner, L.N. Pfeiffer, K.W. West, Measurements of quasiparticle tunneling in the $v$=5/2 fractional quantum Hall state, *Phys. Rev. B*, **85**(16), 165321, (2012).
122. H. Fu, P. Wang, P. Shan, L. Xiong, L.N. Pfeiffer, K. West, M.A. Kastner, X. Lin, Competing $v$ = 5/2 fractional quantum Hall states in confined geometry, *Proc. Natl. Acad. Sci.*, **113**(44), 12386-12390, (2016).

123. A. Stern, B.I. Halperin, Proposed experiments to probe the non-Abelian $v = 5/2$ quantum Hall state, *Phys. Rev. Lett.*, **96**(1), 016802, (2006).
124. P. Bonderson, A. Kitaev, K. Shtengel, Detecting non-Abelian statistics in the $v = 5/2$ fractional quantum hall state, *Phys. Rev. Lett.*, **96**(1), 016803, (2006).
125. A. Stern, B. Rosenow, R. Ilan, B.I. Halperin, Interference, Coulomb blockade, and the identification of non-Abelian quantum Hall states, *Phys. Rev. B*, **82**(8), 085321, (2010).
126. R.L. Willett, K. Shtengel, C. Nayak, L.N. Pfeiffer, Y.J. Chung, M.L. Peabody, K.W. Baldwin, K.W. West, *Interference measurements of non-Abelian e/4 & Abelian e/2 quasiparticle braiding*, arXiv:1905.10248, (2019).
127. N. Jiang, X. Wan, Recent progress on the non-Abelian $v= 5/2$ quantum Hall state, *AAPPS Bulletin* **29**(1), 58-64, (2019).
128. D.E. Feldman, Y. Gefen, A. Kitaev, K.T. Law, A. Stern, Shot noise in an anyonic Mach-Zehnder interferometer, *Phys. Rev. B*, **76**(8), 085333, (2007).
129. D.E. Feldman, A. Kitaev, Detecting non-Abelian statistics with an electronic Mach-Zehnder interferometer, *Phys. Rev. Lett.*, **97**(18), 186803, (2006).
130. V.V. Ponomarenko, D.V. Averin, Mach-Zehnder interferometer in the fractional quantum Hall regime, *Phys. Rev. Lett.*, **99**(6), 066803, (2007).
131. C.J. Wang, D.E. Feldman, Identification of 331 quantum Hall states with Mach-Zehnder interferometry, *Phys. Rev. B*, **82**(16), 165314, (2010).
132. G. Campagnano, O. Zilberberg, I.V. Gornyi, D.E. Feldman, A.C. Potter, Y. Gefen, Hanbury Brown-Twiss interference of anyons, *Phys. Rev. Lett.*, **109**(10), 106802, (2012).
133. G. Yang, Probing the $v=5/2$ quantum Hall state with electronic Mach-Zehnder interferometry, *Phys. Rev. B*, **91**(11), 115109, (2015).
134. J.B. Pendry, Quantum limits to the flow of Information and entropy, *J. Phys. a-Math. Gen.*, **16**(10), 2161-2171, (1983).
135. C.L. Kane, M.P.A. Fisher, Quantized thermal transport in the fractional quantum Hall effect, *Phys. Rev. B*, **55**(23), 15832-15837, (1997).
136. A. Cappelli, M. Huerta, G.R. Zemba, Thermal transport in chiral conformal theories and hierarchical quantum Hall states, *Nucl. Phys. B*, **636**(3), 568-582, (2002).
137. L.G.C. Rego, G. Kirczenow, Fractional exclusion statistics and the universal quantum of thermal conductance: A unifying approach, *Phys. Rev. B*, **59**(20), 13080-13086, (1999).
138. S. Jezouin, F.D. Parmentier, A. Anthore, U. Gennser, A. Cavanna, Y. Jin, F. Pierre, Quantum limit of heat flow across a single electronic channel, *Science*, **342**(6158), 601-604, (2013).
139. M. Banerjee, M. Heiblum, A. Rosenblatt, Y. Oreg, D.E. Feldman, A. Stern, V. Umansky, Observed quantization of anyonic heat flow, *Nature*, **545**(7652), 75-79, (2017).
140. M. Banerjee, M. Heiblum, V. Umansky, D.E. Feldman, Y. Oreg, A. Stern, Observation of half-integer thermal Hall conductance, *Nature*, **559**(7713), 205-210, (2018).
141. F.C. Wellstood, C. Urbina, J. Clarke, Hot-electron effects in metals, *Phys. Rev. B*, **49**(9), 5942-5955, (1994).
142. V. Umansky, M. Heiblum, MBE growth of high-mobility 2DEG, *Book MBE growth of high-mobility 2DEG*, (Elsevier Science BV, Netherlands, 2013).
143. V. Umansky, M. Heiblum, Y. Levinson, J. Smet, J. Nubler, M. Dolev, MBE growth of ultra-low disorder 2DEG with mobility exceeding 35 x$10^6$ cm$^2$/V s, *J. Cryst. Growth*, **311**(7), 1658-1661, (2009).

144. K.K.W. Ma, D.E. Feldman, Partial equilibration of integer and fractional edge channels in the thermal quantum Hall effect, *Phys. Rev. B*, **99**(8), 085309, (2019).
145. S.H. Simon, Interpretation of thermal conductance of the $v = 5/2$ edge, *Phys. Rev. B*, **97**(12), 121406, (2018).
146. D.E. Feldman, Comment on "Interpretation of thermal conductance of the $v$=5/2 edge", *Phys. Rev. B*, **98**(16), 167401, (2018).
147. S.H. Simon, B. Rosenow, *Partial equilibration of the anti-Pfaffian edge due to Majorana disorder*, arXiv:1906.05294, (2019).
148. L. Pfeiffer, K.W. West, The role of MBE in recent quantum Hall effect physics discoveries, *Physica E*, **20**(1-2), 57-64, (2003).
149. M. Samani, A.V. Rossokhaty, E. Sajadi, S. Luscher, J.A. Folk, J.D. Watson, G.C. Gardner, M.J. Manfra, Low-temperature illumination and annealing of ultrahigh quality quantum wells, *Phys. Rev. B*, **90**(12), 121405, (2014).
150. P.V. Lin, F.E. Camino, V.J. Goldman, Electron interferometry in the quantum Hall regime: Aharonov-Bohm effect of interacting electrons, *Phys. Rev. B*, **80**(12), (2009).
151. L. Antonić, J. Vučičević, M.V. Milovanović, Paired states at 5/2 : Particle-hole Pfaffian and particle-hole symmetry breaking, *Phys. Rev. B*, **98**(11), 115107, (2018).
152. C. Wang, A. Vishwanath, B.I. Halperin, Topological order from disorder and the quantized Hall thermal metal: Possible applications to the $v$=5/2 state, *Phys. Rev. B*, **98**(4), 045112, (2018).
153. D.F. Mross, Y. Oreg, A. Stern, G. Margalit, M. Heiblum, Theory of disorder-induced half-integer thermal Hall conductance, *Phys. Rev. Lett.*, **121**(2), 026801, (2018).
154. W. Zhu, D. N. Sheng, *Disorder-driven transition and intermediate phase for $v = 5/2$ fractional quantum Hall effect*, arXiv:1809.04776, (2018).